\documentclass[12pt]{article}
\usepackage{jcappub, amssymb, amsmath,bbm,pifont}
\usepackage{color}

\newcommand{\A}{\mathcal{A}}

\def\ba{\begin{eqnarray}}
\def\ea{\end{eqnarray}}
\def\mpl{M_{\rm Pl}}

\def\d{\mathrm{d}}
\def\S{\Sigma}

\def\L{\mathcal{L}}
\def\({\left(}
\def\){\right)}
\def\ie{{\it i.e. }}

\def\nn{\nonumber}
\def\p{\partial}
\def\mn{_{\mu \nu}}

\def\p{\partial}

\def\<{\langle}
\def\>{\rangle}

\def\A{\mathcal{A}}

\title{\LARGE Unitary NEC violation in $P(X)$ cosmologies}
\author[a,b]{Claudia de Rham}
\author[a]{and Scott Melville}
\affiliation[a]{Theoretical Physics, Blackett Laboratory, Imperial College, London, SW7 2AZ, U.K.}
\affiliation[b]{CERCA, Department of Physics, Case Western Reserve University, 10900 Euclid Ave, Cleveland, OH 44106, USA}

\emailAdd{c.de-rham@imperial.ac.uk}
\emailAdd{s.melville16@imperial.ac.uk}

    \abstract{
A non-singular cosmological bounce in the Einstein frame can only take place if the Null Energy Condition (NEC) is violated. We explore situations where a single scalar field drives the NEC violation and derive the constraints imposed by demanding tree level unitarity on a cosmological background. We then focus on the explicit constraints that arise in $P(X)$ theories and show that constraints from perturbative unitarity make it impossible for the NEC violation to occur within the region of validity of the effective field theory without also involving irrelevant operators that arise at a higher scale that would enter from integrating out more massive degrees of freedom.
Within the context of $P(X)$ theories we show that including such operators allows for a bounce that does not manifestly violate tree level unitarity, but at the price of either imposing a shift symmetry or involving technically unnatural small operator coefficients within the low-energy effective field theory.}


\begin{document}
\maketitle


\tableofcontents

\section{Introduction}

%
Describing the earliest moments of our Universe remains one of the greatest challenges of physics. The singularity theorem states that if the Universe is described by General Relativity (GR) with an FLRW metric and matter that respects the null energy condition, then extrapolating backwards in time, our present understanding must break down and `new physics' has to become important. This can happen in one of two ways:
(i) Either the Hubble parameter reaches Planckian values  at very early times, and the effects of quantum gravity become important (`Big Bang singularity'),
(ii) or the Null Energy Condition (NEC) is violated and the Universe underwent a cosmological bounce.
While theories of quantum gravity are still in development, scenario (ii) may be amenable to current QFT techniques. A great deal of recent work has focused on constructing an early Universe cosmology which takes advantage of (ii) in order to have a non-singular bounce, some of which can even provide an  alternative to inflation (see for instance \cite{Barrow:1993hp,Gasperini:1992em,Khoury:2001wf,Steinhardt:2001st, Khoury:2001bz,ArkaniHamed:2003uy,Battefeld:2014uga,Peter:2006hx, Buchbinder:2007ad,Peter:2008qz,Novello:2008ra,lehners_ekpyrotic_2010,Ijjas:2015zma,brandenberger_bouncing_2016,Peter:2016kan,Cai:2016thi,Cai:2017tku}).\\

In addition to modeling the very early Universe, violating the NEC in a stable, unitary way can be useful in other contexts \cite{Visser:1999de}. It has been proposed as a mechanism for quintessence \cite{Schulz:2001yx}, solving the cosmological constant problem \cite{Abbott:1984qf, ArkaniHamed:2003uy, Alberte:2016izw}, and is also required in theories with traversable wormholes \cite{Hochberg:1997wp, Hochberg:1998ha, Hochberg:1998ii}.\\

Whether NEC violation can ever be stable (free of ghosts and gradient instabilities) has been the subject of much discussion \cite{Holdom:2004yx, Cline:2003gs, Dubovsky:2005xd, Creminelli:2006xe}. In \cite{Deffayet:2010qz}, it was for instance shown that within the context of `Kinetic Gravity Braiding,' (see also \cite{Pujolas:2011he}) one can reach a NEC violating phase while remaining free of ghosts and gradient instabilities.
In the context of the most general single scalar degree of freedom  coupled to gravity (so-called `Horndeski gravity' \cite{Horndeski:1974wa} or `Generalized Galileons' \cite{Deffayet:2009mn}), it is known that there is always a ghost or gradient instability in any bouncing solution as was first pointed out in \cite{Easson:2011zy} and further in \cite{Rubakov:2013kaa, Libanov:2016kfc, Kobayashi:2016xpl,Akama:2017jsa}. However this instability can be made to occur long before or after the NEC violation as was already constructed in the context of G-bounces in \cite{Easson:2011zy} and more recently in a cubic Galileon bounce in \cite{Ijjas:2016tpn}, or in some cases removed by imposing particular asymptotic conditions \cite{Ijjas:2016vtq}. \\

Within the context of single scalar field $P(\Phi, X)$ theories (denoted as simply $P(X)$ in this work unless we wish to emphasize the distinction), it is well-known that the bounce, or the onset of the NEC violation, is necessarily linked with a vanishing speed of sound, and potentially classical instabilities. A way out is to include higher derivative operators in the effective field theory (EFT) which may naturally capture the effect of high energy degrees of freedom without needing to commit to a particular UV completion of the scalar effective field theory considered. Such additional higher derivative operators were previously used in an attempt to regulate the sound speed in the simplest scalar field $P(X)$ theories \cite{ArkaniHamed:2003uy}, for example by adding higher spatial derivatives and arranging for the instability to be much slower than the Hubble rate \cite{Creminelli:2006xe}. \\

In addition to potential classical stability issues linked with a vanishing speed of sound in a relativistic field theory, we emphasize here that there are strong coupling issues associated with it which signal the breakdown of the quantum effective field theory. This can be seen intuitively by noting that the non-gaussianities typically scale as $f_{\rm NL}\sim c_s^{-2}$. A vanishing of the sound speed therefore directly implies that the effect from the higher order operators dominates and hence a breakdown of perturbative unitarity.  To make this statement more concrete and quantify it,
we consider the scattering amplitude of high frequency fluctuations on a given bouncing background and determine the energy scale at which perturbative unitarity is violated. Strong coupling effects from cubic operators in $P(X)$ cosmological bounces were previously discussed in  \cite{Koehn:2015vvy}. Here we explore all the operators that enter the EFT (cubic and beyond) and fully investigate the effect that a small sound speed has on the validity of the EFT. In particular we show that a vanishing sound speed inexorably leads to a vanishing of the strong coupling scale, which would be unacceptable, which agrees with the well-known results of \cite{Vikman:2004dc,Easson:2016klq}. Our conclusion on the fate of the strong coupling issue in any {\it pure}\footnote{Here by pure we mean without the inclusion of high energy irrelevant operators.} $P(X)$  cosmological bounce (including a $P(\Phi,X)$ bounce) therefore departs  from  some other previous analysis, but agrees with \cite{Vikman:2004dc,Easson:2016klq}.  \\

However, by including irrelevant operators that enter from high energy effects, we show that it is possible to restore perturbative unitarity throughout the bounce in $P(X)$ theories. This was for instance proposed in \cite{Creminelli:2006xe, Buchbinder:2007ad} to avoid any gradient instabilities. We show explicitly here how these operators affect the strong coupling scale.  In particular we show that for these high energy irrelevant operators to `save' perturbative unitarity their associated scale needs to be chosen carefully: It must be low enough to restore unitarity, but high enough so as to decouple the low-energy effective field theory from specific high-energy completions.\\

In what follows we shall focus our attention on the requirements set by perturbative unitarity (and particularly tree-level unitarity) within the vicinity of a NEC violating region, so as to determine whether a classical NEC violation can be trusted. We will provide an example where this NEC violation leads to a cosmological bounce, but do not incorporate this bounce within a full cosmological scenario. The model considered here does not attempt to circumvent the no-go mentioned in \cite{Rubakov:2013kaa, Libanov:2016kfc, Kobayashi:2016xpl,Akama:2017jsa}. Rather, the approach of this manuscript is that our $P(X)$ theory ought to successfully capture the duration of the bounce, but that this low energy effective field theory will likely break down (and new physics ought to be included) if followed sufficiently far in the past (well before the bounce or the start of the NEC violating region). \\

With this approach in mind, we will begin with a general discussion of EFT consistency conditions in section~\ref{sec:eft}, which makes few assumptions on the details of the scalar field theory considered. This involves considering a precisely defined {\it decoupling limit} of the gravity/scalar effective field theory, and establishing the constraints arising from the requirement of a (at least perturbatively) unitary $S$ matrix. We then turn our attention in section~\ref{sec:HighEnergy} to the high energy effects that can help regulate the classical gradient instabilities as well as the violation of unitarity at the onset of a NEC violation. Keeping track of those effects, we derive the full bound, bearing in mind that the high energy effects have to occur at a sufficiently low scale to restore unitarity, but at a sufficiently large scale to decouple the high energy states from the low energy effective field theory. This provides a restricted window of possibility.  After having derived the formal requirements set by perturbative unitarity, we focus our attention in section~\ref{sec:PX} to $P(X)$ models that violate the NEC. We show that a level of tuning is required for those models which may call into question their naturalness, but show that in principle a violation of the NEC could occur while maintaining classical stability and perturbative unitarity once a particular (albeit technically unnatural) tuning is chosen. We illustrate this result by providing an explicit covariant example of a $P(\Phi,X)$ model that can allow for a bouncing solution while preserving perturbative unitarity.  Our results are then summarized in section~\ref{sec:conc}. In Appendix~\ref{App:GC} we analyze the well-known ghost-condensate model and show that NEC violation in this model cannot preserve unitarity unless the irrelevant operators coming from high energy effects take a particular form that breaks the shift symmetry.

\section{Single Scalar EFT}
\label{sec:eft}

To set the stage, we start by considering the theory of a single scalar field $\Phi$  coupled to a metric $g\mn$, with no mention of any additional high derivative operators. The consistency of this EFT requires absence of gradient and ghost instabilities as well as unitarity (in the sense that any $n$-point scattering amplitude should satisfy the optical theorem).

\subsection{Decoupling Limit}
\label{sec:dL}
Throughout this manuscript we will focus on a {\it decoupling limit} of the full theory, which is designed to focus on the leading interactions which determine the strong coupling scale of the theory, \ie the scale at which perturbative unitarity is broken. The existence of this decoupling limit comes from the assumption of a hierarchy of scales $\Lambda \ll \mpl$ where $\Lambda$ is the typical interaction scale for the scalar field. Any stability and unitarity bounds determined in that limit represent necessary conditions which must be satisfied by the full theory. In other words, it is sufficient to use the decoupling limit to infer the strong coupling physics, even if not the full cosmological predictions (e.g. power spectrum, bi-spectrum etc.).  Crucially, this limit will continue to allow for bouncing solutions which temporarily violate the null energy condition.
\\

To derive the decoupling limit, we will begin with an action of the form
\ba
\label{eq:Soriginal}
S[g\mn, \Phi]=\int \d^4 x \sqrt{-g}\(\frac{\mpl^2}{2}R +\L(g, \Phi)\)\,.
\ea
We then expand this action around the cosmological background describing the bouncing solution ($\bar \gamma\mn$ and $\phi$), defining the perturbations in an {\it arbitrary gauge} ($h\mn$ and $\varphi$) via
\ba
g\mn=\bar \gamma \mn + h\mn\qquad {\rm and }\qquad
\Phi= \phi+ \varphi\,.
\ea
The physical action for the perturbations is (schematically)
\ba
S[h\mn,\varphi]=  \int \d^4 x \sqrt{-\bar \gamma} \Bigg(- \mpl^2 \, \bar{\mathcal{B}}(\bar \gamma)(\p h)^2 - \bar R(\bar \gamma) h^2 + \mpl^2 f_1(\bar\gamma) h (\p h)^2 +\cdots \\
-  Z(\bar \gamma,  \phi)(\p \varphi)^2 -  m^2 (\bar \gamma,  \phi) \varphi^2 + f_2(\bar \gamma, \phi) \varphi (\p \varphi)^2+ f_3(\bar \gamma, \phi) \varphi^3  + \cdots\nn\\
+ f_4(\bar \gamma,  \phi) h \p^2 \varphi +  f_5(\bar \gamma,  \phi) h  \varphi + f_6(\bar \gamma,  \phi) h^2 \p^2 \varphi + \cdots \nn
\Bigg)\,,
\ea
where the last line represents all the potential mixing between $h$ and $\varphi$. All the functions $\bar{\mathcal{B}}$, $\bar R$, $Z$ and $f_{i}$ depend on the background (and of course carry indices), however their exact expressions are irrelevant for the rest of the scaling argument.  The scale of the background naturally enters all these functions---for instance, $\bar R(\bar \gamma)$ is related to the background curvature. In principle, the linear kinetic mixing between $h$ and $\varphi$ could be taken care of by performing an appropriate field redefinition and absorbing the function $f_3/\mpl^2$ in the expression of $Z$. In practice however such a shift is irrelevant in the decoupling limit we will consider below as it is suppressed by the Planck scale. \\

Our working assumption, will be that the scalar field contains interactions at the scale $\Lambda$, such that $\L(g, \Phi ) \sim \Lambda^4$, and that there is a large hierarchy $\Lambda \ll \mpl$. If this is the case then we typically expect $\dot H \sim \Lambda^4/\mpl^2$. The leading interactions which determine the strong coupling physics will be determined principally by the scalar field, with those that arise from the mixing with gravity being $\mpl$ suppressed. As such we may take a decoupling limit $\mpl \rightarrow \infty$, keeping $\Lambda$ fixed. In the explicit bouncing solutions we construct in section~\ref{Explicitmodel}, the time scale for the null energy violation is set by the scale of the scalar interactions $\Delta t \sim \Lambda^{-1}$. This in turn implies an even stronger suppression for the Hubble rate $H \sim \Lambda^3/\mpl^2$, since $H \sim \dot H \Lambda^{-1}$. This is borne out by the explicit solutions \ref{Explicitmodel}.  \\

The first step in dealing with this effective field theory of $h\mn$ and $\varphi$ is to properly canonically normalize the fields.  In this case, the appropriate canonical normalization of the field is\footnote{In practice the tensors $\bar{\mathcal B}$ and $Z$ are usually not conformal with respect to the background metric and the rescaling should be taken with slightly more care as will be performed in section~\ref{sec:scalarFLRW}, however those subtleties do not affect the essence of the decoupling limit. }
\ba
\label{eq:normalization}
\tilde h\mn \sim \frac{1}{\sqrt{\bar{\mathcal{B}}} \mpl} h\mn \qquad {\rm and} \qquad \tilde \varphi \sim \frac{1}{\sqrt{Z}} \varphi\,.
\ea
and the action is (again symbolically)
\ba
S[\tilde h\mn,\tilde \varphi]=  \int \d^4 x \sqrt{-\bar \gamma} \Bigg( -(\p \tilde h)^2 - \frac{\bar R}{\bar{\mathcal{B}}} {\tilde h}^2   -(\p \tilde \varphi)^2 -  \frac{m^2}{Z} \tilde \varphi^2 + \frac{f_2}{Z^{3/2}} \tilde \varphi (\p \tilde \varphi)^2 + \frac{f_3}{Z^{3/2}} \tilde \varphi^3  + \cdots\quad \\
+ \frac{1}{\mpl} \frac{f_1}{\bar{\mathcal{B}}^{3/2}} \tilde h (\p \tilde h)^2
+ \frac{f_4}{\mpl \sqrt{\bar{\mathcal{B}} Z}}\tilde h \p^2 \tilde \varphi +  \frac{f_5}{\mpl \sqrt{\bar{\mathcal{B}} Z}} \tilde h  \tilde \varphi + \frac{f_6}{\mpl^2 \bar{\mathcal{B}}\sqrt{Z}} \tilde h^2 \p^2 \tilde \varphi + \cdots \nn
\Bigg)\,.
\ea
Then taking a limit where $\mpl\to \infty$, while maintaining the scales that enter the scalar field background fixed, we see that all of the whole second line becomes unimportant, and the scalar field fluctuations $\tilde \varphi$ entirely decouple from the metric fluctuations (which becomes a trivial free theory). A significant virtue of this decoupling limit, is that the gauge degrees of freedom remain in $h$ and decouple. This procedure is thus {\it insensitive to any gauge issues}. For this reason it will be unnecessary to work with the comoving curvature perturbation $\zeta$, or similar gauge invariant variables (this point is discussed in more detail at the end of this subsection).
\\

The relevant effective action in this limit is hence solely that of the scalar field fluctuations
\ba
S_{\rm dec}= \int \d^4 x \sqrt{-\bar \gamma} \(- (\p \tilde \varphi)^2 -  \frac{m^2}{Z} \tilde \varphi^2 + \frac{f_2}{Z^{3/2}}\tilde \varphi (\p \tilde \varphi)^2+ \frac{f_3}{Z^{3/2}} \tilde \varphi^3 + \cdots \)\,,
\ea
where the ellipses carry operators to all orders in $\varphi$ (and potentially $\p \varphi$ and even $\p^2 \varphi$ and higher order in derivatives). Although gravity has decoupled, this is not the same as the scalar theory on Minkowski spacetime. Information about the background is carried through the process, and we are effectively looking at a scalar field on curved background. In the explicit solutions given in section~\ref{Explicitmodel}, $\phi(t)$ remains finite in the limit $\mpl \rightarrow \infty$, with a time dependence at the scale $\Delta t \sim 1/\Lambda$. In other words the scalar field is varying at a time scale $1/\Lambda$ which is much faster than naive background scale $1/H$, similarly $\dot H /H \sim  \Lambda \gg H$.
\\

Before proceeding, it is worth noting that a decoupling limit of this form is not always appropriate, for example in slow roll inflation, the flatness of the potential means that the interactions coming from the mixing with gravity are actually the dominant ones, and it would be incorrect to perform the above limit. Another example would be if the interactions with the metric were made artificially large, for instance in the following example,
    \ba
    S=\int \d^4 x \sqrt{-g}\(\mpl^2 R  + \frac{\mpl}{\Lambda}R  \Phi^2 -\frac 12 (\p \Phi)^2+\frac{1}{\Lambda^4} \Phi^4\)\,,
    \ea
 where $\Lambda \ll \mpl$. In this case the mixing term between the metric and the scalar field fluctuations ought to be taken into account for background configurations with $\phi \sim \Lambda$. Of course in this case, we can simply remedy this issue by first going to Einstein frame, which automatically accounts for the scalar/gravity mixing. When dealing with generic scalar-tensor theories there is not necessarily a covariant definition of Einstein frame, however the appropriate procedure would simply be to first go to the relevant Einstein frame at the perturbed level about the cosmological background and then perform the appropriate decoupling limit. In the present case, the decoupling limit we have defined is justified by the existence of a large class of explicit solutions, which we discuss in section~\ref{Explicitmodel}, which have the property that $H \sim \Lambda^3/\mpl^2$ and $\dot H \sim \Lambda^4/\mpl^2$. \\

We emphasize that our concern here is about the consistency of the effective action, and the scale at which perturbative unitarity breaks down. This is a very different question than that of say, the precise form of the power spectrum or bispectrum. For precise questions of this sort, performing a decoupling limit and focusing on the scalar field effective action would not provide an accurate answer, however for the question of perturbative unitarity we wish to address in this manuscript, focusing on the scalar field decoupling limit on the appropriate background gives necessary conditions which must be satisfied.\\

\paragraph{Gauge issues:} Before analyzing this decoupling limit we briefly comment on the gauge issues that have been highlighted in the literature (see for instance Ref.~\cite{Koehn:2015vvy} for a recent discussion). In particular if we first chose a gauge where the scalar part of  $h_{ij}$ vanishes we would find that the constraints for the shift and the lapse would impose them to scale as $(\mpl^2 H)^{-1}$. The resulting Lagrangian (before taking any decoupling limit) would then involve inverse powers of $\mpl^2 H$. Since in the explicit solution we provide in section~\ref{Explicitmodel} $H\sim \Lambda^3/\mpl^2$, such a behaviour, if generic, would invalidate the decoupling limit arguments. Moreover there would seem to be an `apparent' singularity at the bounce itself (\ie when $H=0$). This should come as no surprise since it is  impossible to fix the gauge $h_{ij}^{\rm scalar}=0$ when $H=0$, so the previous apparent singularity is simply a gauge artefact and so is the scaling found in that gauge. \\

Switching to comoving gauge however does not help with the previous issue either since in that gauge the kinetic coefficient of the curvature perturbation does itself then scale as $H^{-1}$ which makes it impossible to properly normalize the field. Rather than focusing on any of these two gauges (or any standard `local' gauges), the problem at hand can be entirely dealt with by going for instance to de Donder gauge or any other gauge of that form (or at leading order by going to harmonic gauge as performed in \cite{Battarra:2014tga}). The appropriate way to perform this is to consider the action \eqref{eq:Soriginal} and add the appropriate Fadeev-Popov gauge fixing terms so that they combine with the  Einstein curvature term to give kinetic terms for the metric fluctuation that take the remarkably simple form $\mpl^2 h^{\mu \nu} \bar \Box \(h\mn -\frac 12 h \bar \gamma\mn\)$ in addition to higher order fluctuations, (and where $\bar \Box $ is the d'Alembertian with respect to the background metric $\bar \gamma\mn$).  In this language there is no constraint to solve for since the gauge fixing terms are precisely there to break gauge invariance. The absence of constraints ensures that at no point one would need to perform an inversion of the Hubble parameter and  the appropriate canonical normalization follows the same behaviour as in \eqref{eq:normalization} (where $\mathcal{B}$ is manifestly finite and is trivial in the flat space limit). The breaking of gauge invariance from the gauge fixing terms comes at the price of including other spurious degrees of freedom  but it is well understood how to introduce the  Fadeev-Popov ghosts to deal with those and they only contribute to loops. All the tree-level amplitudes computed with  these de Donder gauge fixing terms are therefore the same as that of the original theory. In this formulation the decoupling limit can therefore be taken precisely as discussed previously. For the questions we are interested in (namely the size of the strong coupling scale, whether or not the theory preserves tree-level unitarity, etc...), we can therefore safely perform this decoupling limit and work with the low-energy effective field theory for the scalar field $\varphi$ on the cosmological background. \\

In the case of the specific example that will be provided in section~\ref{Explicitmodel}, we can compute (at least to a given order) the corrections that arise beyond  the decoupling limit. By performing a complementary  analysis to that described above, and keeping track of the corrections that arise due to the mixing with the metric, we have checked that we obtain a correction to the effective mass for the scalar field which is suppressed by six orders of magnitude, which is precisely what one would have expected in our example.

\subsection{Scalar on FLRW}
\label{sec:scalarFLRW}

Since the main interest of this manuscript is cosmology  (and the potential description of cosmological bounces), the following analysis will take place on a flat FLRW background with scale factor $a (t)$. The effective metric for the scalar field fluctuations then takes the form
\ba
\label{eq:Z}
Z^{\mu\nu} \p_\mu \varphi \p_\nu \varphi = - A(t) \dot \varphi^2 + \frac{B(t)}{a^2} (\p_i \varphi)^2\,,
\ea
where $A$ and $B$ depend on the background behaviour (here and in what follows dots represent the physical time derivatives and $\p_i$ designate spatial gradients).
The corresponding sound speed is $c_s^2=\frac{B}{A}$.
The absence of a ghost (in the scalar sector) then implies $A>0$, while the absence of gradient instabilities implies $B>0$. \\

For the rest of the analysis we will assume that we are within the WKB regime, meaning that we consider  frequencies well above the scale set by the background time variation, $E_{\rm back}$ with
\ba
\label{eq:Ebackground2}
E_{\rm back} = {\rm Max}\(H, \sqrt{\dot H}, m,  \dot \phi/ \phi, \ddot \phi / \dot \phi, \cdots \)\,.
\ea
 Indeed to probe unitarity violation we are interested in analyzing the interactions of modes close to the strong coupling scale, which should be much larger than the background scale for the effective field theory to make sense.  In the explicit models given later, the scale of variation of the background solution will be $1/\Lambda$, \ie $E_{\rm back} \sim \Lambda$, and so to ensure the validity of the WKB regime we require $\omega\gg \Lambda$.
\\

\paragraph{Canonical normalization:}

Before going through the formal unitarity bounds arising from the optical theorem, it is useful to first estimate the strong coupling scale by simply analyzing the scale of the operators present in the scalar field theory (on the FLRW background). Since Lorentz invariance is {\it spontaneously} broken by the FLRW background, the effective metric $Z^{\mu\nu}$ is not conformally flat, so before canonically normalizing the field, it is useful to perform the following coordinate rescaling\footnote{As we shall see, as soon as $B$, or the speed of sound approaches small enough (positive) values, the whole EFT runs out of control and the classical background is not to be trusted, therefore there is no sense in which $B$ or the speed of sound actually ever vanishes, much less became negative in the first place. The rescaling performed in this section is thus well defined.}
\ba
\tilde t = \int c_s (t) \d t \qquad {\rm and }\qquad \tilde{x}^i = x^i
\ea
so that in this system of coordinates the effective metric is conformally flat,
\ba
\int \d t \d^3 x \, a^3\(  -\frac 12 Z^{\mu\nu} \p_\mu \varphi \p_\nu \varphi \)= \int \d \tilde t \d^3 \tilde x \, a^3 \frac{B}{2c_s}
\(\(\frac{\p \varphi}{\p \tilde t}\)^2-\frac{1}{a^2}\(\frac{\p \varphi}{\p \tilde{x}^i}\)^2\)\,,
\ea
and we can simply canonically normalize the field by setting
\ba
\varphi=\sqrt{\frac{c_s}{B}} \, \tilde \varphi =\(A B\)^{-1/4}\,  \tilde \varphi \,,
\ea
(the derivatives of $AB$ are then simply absorbed into the mass term).\\

\paragraph{Irrelevant Operators:}
Now consider an EFT on the FLRW background that contains an irrelevant operator of the form
\ba
\label{eq:OperatorsNML}
S_{NML} = \int \d t \d^3 x  \frac{a^{3-2L}}{\Lambda_{NML}^{N+2M+4L-4}} \varphi^N \dot \varphi^M (\p_i \varphi)^{2L}\,,
\ea
where  $N$, $M$ and $L$ are arbitrary positive integers with $N+2M+4L> 4$ and the scale $\Lambda_{NML}$ depends on the background and the particular theory one is dealing with.
 Then in terms of the rescaled coordinates and the properly canonically normalized field, this interaction is
\ba
\label{eq:OperatorsNML2}
S_{NML} = \int \d \tilde t \d^3 \tilde x \frac{ a^{3-2L}}{\tilde \mu_{NML}^{N+2M+4L-4}} (\tilde\p_i \tilde \varphi)^{2L} \sum_{j=0}^M \(\frac{\p_{t}(AB)}{c_s AB}\)^j \tilde\varphi^{N+j} \(\p_{\tilde t}\tilde \varphi\)^{M-j} \,,\qquad
\ea
where we have ignored signs and other order one and combinatory numbers. The scale $\tilde \mu$ is given by
\ba
\tilde \mu_{NML}= A^{\frac{N+3M+2L-2}{4(N+2M+4L-4)}} B^{\frac{N-M+2L+2}{4(N+2M+4L-4)}} \Lambda_{NML} \,.
\ea
Now when rescaling back to the original coordinates, as an energy scale we have
\ba
\mu_{NML} = \frac{\p \tilde t}{\p t} \, \tilde \mu_{NML}= c_s\,  \tilde{\mu}_{NML}\,,
\ea
and so the scale that enters these interactions is
\ba
\label{eq:muNML}
\mu_{NML}=A^{n_A} B^{n_B} \Lambda_{NML} \,,
\ea
with
\ba
\label{eq:nAnB}
n_A = - \frac{N+M+6L-6}{4(N+2M+4L-4)} \quad {\rm and}\quad  n_B =  \frac{3N+3M+10L-6}{4(N+2M+4L-4)}\,.
\ea
We therefore see that in a situation where $A$ becomes parametrically large or $B$ is parametrically small, for a fixed scale $\Lambda_{NML}$ there are typically operators that enter at a parametrically small scale $\mu_{NML}\ll \Lambda_{NML}$ and therefore spoil the validity of the effective field theory at a low scale. \\

For these arguments to be valid one should have $\frac{\p_{t}(AB)}{AB}\sim E_{\rm back} \lesssim {\rm Min}(\mu_{NML})$, (where ${\rm Min}(\mu_{NML})$ is the lowest of all the possible scales derived in \eqref{eq:muNML} for any non-negative integers $N,M,L$ with $N+2M+4L>4$). This allows us to ignore the contributions from $\frac{\p_{t}(AB)}{c_s AB}$ in (\ref{eq:OperatorsNML2}). If $\frac{\p_{t}(AB)}{ AB} \sim E_{\rm back}$ was larger than any of those scales $\mu_{NML}$, then by definition the background would be varying faster than the time scale set by the strong coupling scale of the effective theory and it would no longer be possible to trust the validity of the background solution.

\paragraph{Marginal and Relevant Operators:}

For the relevant $\Lambda_{300} / 3!  \varphi^3$ and marginal $\lambda/4! \varphi^4$ operators, the previous rescaling can also be performed and leads to
\ba
\label{eq:mu300}
\mu_{300} &=& \frac{\Lambda_{300}}{(AB)^{3/4}}\\
\mu_{400} &=& \frac{\lambda}{(AB^3)^{1/2}}\,,
\label{eq:mu400}
\ea
where of course  $\mu_{400}$ does not represent an actual scale but rather a  dimensionless coupling constant.
Remaining in the perturbative regime requires the dimensionless coupling constant to be $\mu_{400} \ll \mathcal{O}(16\pi)$ and the scale of the marginal operator to be $\mu_{300}\lesssim E_{\rm back}$.  However since these operators are renormalizable, we can deal with them in the strong coupling regime and we therefore do not necessarily need to impose that these couplings are small to preserve unitarity.

\subsection{Optical theorem}
\label{sec:Optical1}

The previous section provided a generic scaling argument to determine the typical interaction scale of an operator on a FLRW background. We now provide more substance to this argument by showing how it relates to the optical theorem by computing a  precise scattering amplitude.  \\

Going back to the effective metric \eqref{eq:Z} in its original formulation on FLRW (before any canonical normalization), in the WKB regime the quantized modes can be decomposed in the following way
\ba
\label{eq:modesphi}
\hat \varphi(t,x^i)=\int \frac{\d^3 k_i}{(2\pi)^3 a^3 }\frac{1}{\sqrt{\mathcal{N} (k) }}\(\hat a^\dagger(k_i)e^{i (k \int \frac{c_s(t)}{a(t)} \d t - k_i x^i)}
+ \hat a(k_i)e^{-i (k \int \frac{c_s(t)}{a(t)} \d t - k_i x^i)} \)\,.
\ea
To derive the normalization  $\mathcal{N}(k)$ we go back to the well-known definition of the Klein-Gordon norm along a three-dimensional surface $\Sigma$ with unit normal vector $n^\mu$ and induced metric $\gamma_{\mu\nu}$,
\ba
|\hat \varphi|^2=-i \int_{\Sigma}\d^3x\,  \sqrt{\gamma}\,  n^\mu \(\hat \varphi \overleftrightarrow{\p_\mu} \hat \varphi^\dagger\)\,.
\ea
Choosing a spacelike surface $t={\rm const}$, the unit normal vector is $n^\mu = \sqrt{A} \delta^\mu_0$ and the induced volume element is $\sqrt{\gamma}=B^{-3/2}$, leading to the Klein-Gordon norm,
\ba
|\hat \varphi|^2=-i \frac{\sqrt{A}}{B^{3/2}}\int \d^3 x a^3 \(\hat \varphi \overleftrightarrow{\p_t} \hat \varphi^\dagger\)\,.
\ea
We can therefore infer the field normalization,
\ba
\label{eq:Norm1}
\mathcal{N} (k)= \frac{\p F}{\p \omega} = 2 A \omega=2 (AB)^{1/2} \frac{k}{a} \,,
\ea
where we evaluated the mode on-shell, $\omega=c_s k/a$, and
where the function $F$ determines the dispersion relation,
\ba
\label{eq:DispF1}
F=A \omega^2 - B \frac{k^2}{a^2}=0\,.
\ea
In other words the speed of sound is given by $c_s^2=B/A$. The operators $\hat a$ and $\hat a^\dagger$ that appear in \eqref{eq:modesphi} are the creation and annihilation operators for one-particle states of definite momentum which obey the usual commutation relations.  \\

In the high energy WKB regime, it is meaningful to talk about an approximate S-matrix for scattering of quanta, provided it is defined over a time scale shorter than the background variation  $\Delta t \ll 1/E_{\rm back}$. The constraints from unitarity on this approximate S-matrix are many, however it is useful to focus on the special case of $n$ to $n$ scattering, between states of equal momenta, $| i \rangle= | f \rangle = | k_1 ... k_n \rangle $. This gives a unitarity bound for every integer $n \geq 2$,
\ba
2 {\rm Im} \(\langle i | T | i \rangle\) = \sum_N \langle i | T^\dagger | q_1 ... q_N \rangle \langle q_1... q_N | T | i \rangle \geq \left| \langle q_1...q_n | T | i \rangle \right|^2\,,
\ea
so that the scattering amplitudes $\A_{2n}$ (with momentum conserving delta function removed) should satisfy the following relation
\ba
 2   \text{Im} \; \A_{2n} ( k_j ; k_j  )
&\geq&  \int \frac{\d^3 q_1}{(2 \pi)^3 \mathcal{N}(q_1) } \cdots \frac{\d^3 q_n}{(2 \pi)^3 \mathcal{N}(q_n) } \;\;   | \A_{2n} (k_j ; q_j) |^2   \nn \\
& \times &\left[ (2 \pi)^4 \delta \(\frac{c^2_s}{a^2} \(\sum k^2_j - \sum q^2_j\) \) \delta^{(3)} \(\sum k_j - \sum q_j \) \right]  \,. \quad  \label{eqn:A2n1}
\ea
The scattering amplitudes may be computed using standard Feynman diagrams using the WKB form for the propagator for scattering at time intervals shorter than $\Delta t \ll 1/E_{\rm back}$.\\

\paragraph{4-point function:}
In the case of $n=2$, the two-body phase space factor on the right hand side of \eqref{eqn:A2n1} can be evaluated simply in terms of a center of mass energy, $\sqrt{s}$, and scattering angle, $\theta$, giving the well-known optical theorem for the $2$ to $2$ scattering amplitude $\A_4 (s, \theta)$. For our dispersion relation $\omega^2= c_s^2 k^2/a^2$, a partial wave expansion yields the following bound for every $\ell\ge0$,
\begin{equation}
\Big|  \A_{4,\ell} (s) \Big| \leq 8 \pi^2\,  \frac{\omega}{k/a} A B = 8 \pi^2 \, (AB^3)^{1/2}\,,
\label{eqn:A4boundfull1}
\end{equation}
with
\ba
 \A_{4,\ell} (s) = \int_{-1}^1 \d\cos \theta \, P_{\ell} (\cos \theta) \, \A_4 (s, \theta)\,.
\ea
Now consider a cubic operator  of the form given in \eqref{eq:OperatorsNML} with $N+M+2L=3$ (keeping in mind we are still dealing with an irrelevant operator,  $N+2M+4L> 4$). Taking into account the proper normalization of the propagator, such a cubic operator would lead to a contribution to the 4-point function  scaling as (at fixed angle)
\ba
\label{eq:A4cubic}
\A^{\rm (cubic)}_4 \sim \frac{k^{4L-2}\omega^{2M}}{B\Lambda^{2(N+2M+4L-4)}_{NML}}\sim \frac{A^{2L-1}}{B^{2L}} \(\frac{\sqrt{s}}{\Lambda_{NML}}\)^{2(M+2L-1)}\,,
\ea
(keeping in mind that $N=3-M-2L$). We have ignored any combinatory factor in this estimation since we are mainly after  a scaling argument. However once a  particular amplitude is diagnosed as being potentially problematic the proper factors will be  included.
Now applying the bound \eqref{eqn:A4boundfull1}, we determine that perturbative unitarity gets broken at the scale
\ba
\sqrt{s}\sim A^{\frac{3-4L}{4(M+2L-1)}} B^{\frac{3+4L}{4(M+2L-1)}} \Lambda_{NML}\,,
\ea
which is precisely the scale $\mu_{NML}$ inferred from the simple scaling argument in \eqref{eq:muNML} when $N=3-M-2L$. When dealing with the contribution to the $2-2$ scattering amplitude from the cubic vertex $\varphi^3$, we see that the amplitude \eqref{eq:A4cubic} is dominated instead by IR effects and one should go beyond the tree-level amplitude when  $\Lambda_{300}\gtrsim (AB)^{3/4} E_{\rm back}$, as deduced in \eqref{eq:mu300}. \\

Similarly, we can consider a quartic irrelevant operator  of the form given in \eqref{eq:OperatorsNML} with $N+M+2L=4$ ($N+2M+4L> 4$) which would lead to a contribution to the 4-point function going as (at fixed angle)
\ba
\label{eq:A4quartic}
\A^{\rm (quartic)}_4 \sim \frac{k^{2L}\omega^M}{\Lambda_{NML}^{N+2M+4L-4}} \sim \frac{A^L}{B^L}\(\frac{\sqrt{s}}{\Lambda_{NML}}\)^{M+2L}\,.
\ea
From the perturbative unitarity bound \eqref{eqn:A4boundfull1} we see that such a quartic operator would be responsible for breaking perturbative unitarity at the scale
\ba
\sqrt{s}\sim \Lambda_{NML} A^{\frac{1-2L}{2M+4L}}B^{\frac{3-2L}{2M+4L}}\,,
\ea
 which is again precisely the scale $\mu_{NML}$ derived in \eqref{eq:muNML} for a quartic operator $N=4-M-2L$.
 When dealing with the marginal operator $\lambda \varphi^4/4!$, its contribution to the previous amplitude would simply go as $\lambda$ and remaining in the perturbative regime then requires $\lambda \ll (A B^3)^{1/2}$ which is once again precisely the requirement deduced previously in \eqref{eq:mu400}, although breaking this bound does not yet imply breaking unitarity since that operator is renormalizable. \\

One obvious worry is that cancellations could occur for instance between different cubic operators or between the contributions from $\A^{\rm (cubic)}_4$ and $\A^{\rm (quartic)}_4$. If that were the case, it would simply imply that a field redefinition could be performed to remove such operators (or part of them). For the simple scalar field effective theory we are dealing with this is fortunately relatively straightforward to monitor and provided we are not dealing with an unnecessarily complicated formulation of the effective theory the strong coupling scale would indeed be given by the smallest of the scales $\mu_{NML}$ provided  in \eqref{eq:muNML}. A more complete discussion of this aspect is given in section~\ref{sub:fieldRedef}. \\

\paragraph{$2n$-point function:}
To  complete the perturbative unitarity requirement as directly imposed from the optical theorem, we provide here the bounds from higher $n$-point functions. In that case the $n$-body phase space factor is a complicated integral over several momenta and scattering angles, but those do not affect the overall scaling of the bound and we obtain the following perturbative unitarity bound for the $2n$-point functions,
\begin{equation}
 | \A_{2n} |  \lesssim  A^{(3-n)/2}  B^{3(n-1)/2} s^{2-n} \,,
\label{eqn:A2nboundlow1}
\end{equation}
where we have ignored order one numerical factors.
We can again compare this bound with the strong coupling scale we would infer from an operator \eqref{eq:OperatorsNML} with $N+M+2L=2n$ and we see that we infer precisely the same scaling for the strong coupling scale in terms of $A$ and $B$ as was found in \eqref{eq:muNML}.

\subsection{Field Redefinitions and Redundant Operators}
\label{sub:fieldRedef}

The previous arguments to deduce the strong coupling scale and the breaking of perturbative unitarity implicitly assumed that any operator of the form \eqref{eq:OperatorsNML} leads to a contribution to the tree-level scattering amplitude. There are of course cases where this assumption fails:
\begin{itemize}

\item First, it may possible that the contribution of an operator to a scattering amplitude accidentally cancelled at tree-level. If the cancellation only occurs at tree-level but the operator still contributes to loops with an order of magnitude comparable to what is estimated in for instance \eqref{eq:A4quartic} or \eqref{eq:A4cubic}, then the argument would remain unaffected as the loops will still lead to a violation of perturbative unitarity (even though this may occur at higher order in the loop expansion).

\item Second, the contribution of an operator to {\it any} scattering amplitude may exactly cancel (to all orders in loops). This would simply mean that this operator is actually not present and in a single scalar field theory (where there is no gauge issue) this can only happen if that operator is simply removable by integrations by parts.  This case is of course trivial.

\item Finally the particular contribution of an operator may not cancel by itself but may be canceled by the contributions of other operators. If this cancellation does not occur for scattering amplitudes and to all orders in loops, then the previous arguments on the breaking of perturbative unitarity is effectively unaffected. On the other hand, if this set of operators was simply removable by an appropriate field redefinition without leading to other new higher order operators, then in practice these operators are redundant and superficial to the description of any physics. In practice when analyzing a scalar field effective theory we ought to take care of all such redundancies in the first place.
\end{itemize}
Besides these previous special cases, the arguments presented here  are robust in diagnosing the potentially dangerous operators  and applicable to any scalar field effective field theory which carries operators of the form \eqref{eq:OperatorsNML}. Once a diagnosis has been established one can directly compute the amplitude with the appropriate combinatoric factors to fully determine the fate of unitarity. \\

The main motivation of this work is to apply these bounds to $P(X)$ models that can in principle allow for a violation of the NEC and potentially allow for a cosmological bounce. Within the context of pure $P(X)$ theories (coupled) to gravity, it is well-known that the speed of sound turns negative at the onset of the NEC violation and thus leads to gradient instabilities, this agrees with previous literature such as \cite{Vikman:2004dc,Easson:2016klq}. This can potentially be remedied by considering new irrelevant operators that arise at a higher scale and regulate this instability. For instance considering the irrelevant operator $(\Box \Phi)^2/\Lambda_c^2$ it can allow for a stable bounce \cite{Creminelli:2006xe, Buchbinder:2007ad}. In what follows we therefore explore the effects of such an irrelevant operator on the unitarity bound considered previously, before focusing on $P(X)$ models endowed with such an operator.

\section{Inclusion of high energy effects}
\label{sec:HighEnergy}

We now consider the inclusion of high energy effects that would naturally enter any EFT. For concreteness, we can imagine that the next massive particle enters with a mass $\Lambda_c$, and integrating out such a massive particle leads to an operator for the form\footnote{\label{foot:ghost}Even though the operator considered here carries more derivatives, as we shall see,
by construction, the would be Ostrogradsky ghost associated with it is not present since its mass lies above the cutoff of the EFT. In what follows we take great care in making sure that in the regime of validity of the cosmological bounce, the would-be ghost is not present, see Refs.~\cite{Burgess:2007pt,Creminelli:2010qf,deRham:2014fha} for a discussion of this point.} $(\Box \Phi)^2/\Lambda_c^2$ in the low-energy effective action for $\Phi$. To see this more precisely, we could for instance consider the two-scalar field example coupled to gravity,
\ba
\label{eq:chi}
S[g\mn, \Phi, \chi]=\int \d^4x \sqrt{-g}\(\frac{\mpl^2}{2}R +\L(g, \Phi)  -\frac 12 (\p \chi)^2 -\frac {\Lambda_c^2}{2} \chi^2 +\frac 12 \chi \Box \Phi\)\,,
\ea
where we assume no ghost nor other types of pathologies in $\L(g, \Phi)$. Note that at this level the model is entirely free of any type of  ghost.
If the field $\chi$ is sufficiently massive (\ie the scale $\Lambda_c$ is sufficiently large as compared with the other scales in $\L(g, \Phi)$), the field $\chi$ is frozen and its dynamics decouple, and we can integrate it out. At leading order,
\ba
\chi=-\frac{\Box}{\Box-\Lambda_c^2}\Phi= \frac{\Box \Phi}{\Lambda_c^2}+\cdots\,,
\ea
and we are left with irrelevant operators in the low-energy effective field theory for $\Phi$
\ba
S[g\mn, \Phi]=\int \d^4 x \sqrt{-g}\(\frac{\mpl^2}{2}R +\L(g, \Phi)  +\frac{1}{2\Lambda_c^2} (\Box \Phi)^2+\cdots \) \,,
\ea
where the ellipses designate higher order corrections in $\Box/\Lambda_c^2$. As mentioned in footnote \ref{foot:ghost}, the irrelevant operator we have included $(\Box \Phi)^2$ seems to carry higher derivatives and one could be worried about the associated Ostrogradsky ghost, however in this context, when reaching the scale $\Lambda_c$, it is no longer appropriate to integrate out the dynamics of $\chi$ and one should go back to \eqref{eq:chi} for the appropriate description, which is clearly free of ghosts. \\

In general we could expect the scale $\Lambda_c$ to also be background dependent (e.g. through $\Phi$), and hence to carry a time dependence within this cosmological setup, but within the WKB approximation (keeping in mind that we interested in energy scales much smaller than $\Lambda_c$) we shall ignore any possible time dependence of $\Lambda_c$ for the rest of the manuscript. \\

Since $(\Box \Phi)^2/\Lambda_c^2$ is a higher derivative operator, it leads to inevitable unitarity violation at the scale $\Lambda_c$ (or possibly a background redressed version $\Lambda_c'$). Consequently this sets an upper bound on the cutoff of the effective theory $\Lambda_{\rm cutoff} \le \Lambda_c'$. This cutoff scale is not necessarily the same as the strong coupling scale $\Lambda_s$ which signals the breaking of perturbative unitarity of the theory  (see refs.~\cite{Aydemir:2012nz,deRham:2014wfa} for a clear distinction between these two scales). New physics may not necessarily be required at the scale of the breaking of perturbative unitarity, however the classical background can no longer be trusted in that case. So if a classical violation of the NEC occurred at a scale $E_{\rm back}$ beyond the strong coupling scale, there would be no reason to believe the NEC violation actually took place. Putting this together, then overall consistency of our EFT description requires $E_{\rm back} \ll \Lambda_s \le \Lambda_{\rm cutoff} \le \Lambda_c'$.
\\

\subsection{Dispersion Relation}

At first sight, the inclusion of the operator $(\Box \Phi)^2/\Lambda_c^2$, suppressed by the large scale $\Lambda_c$, would appear to have a negligible effect on the theory. Furthermore if it does have an effect, one would worry that it would be necessary to include the infinite number of higher order operators that enter at the scale $\Lambda_c$, such as $(\Phi \Box^{n+2} \Phi)/\Lambda_c^{2+2n}$. The reason that this is not the case is that in the absence of this operator, for the NEC violating solutions we consider, $c_s$ passes through zero and temporarily becomes negative. The inclusion of the operator $(\Box \Phi)^2/\Lambda_c^2$ then creates a large correction (relative to 0), which must be included, while at the same time higher order operators like $(\Phi \Box^{n+2} \Phi)/\Lambda_c^{2+2n}$ will be negligible. To see how this works, we note that on including the irrelevant operator $(\Box \Phi)^2/\Lambda_c^2$  in the scalar field effective theory on the curved background, the quadratic action for $\varphi$ is then
\ba
\label{eq:S2}
 S^{(2)} = \int \d^4 x \, \frac{a^3}{2} \left(  A \dot \varphi^2 - B \frac{ (\partial_i \varphi )^2}{a^2} -  m^2 \varphi^2 + \frac{1}{ \Lambda_c^2} \left(  \ddot \varphi -  \frac{  \partial_i^2 \varphi}{a^2}  \right)^2 \right)
\ea
where $A,B$ and $m$ are determined by the background behaviour and we still use the notation  $c_s^2=B/A$.
The expression for the function $F$ that provides the equation for the dispersion relation in \eqref{eq:DispF1} is therefore now promoted to
\ba
\label{eq:DispF2}
F=A \omega^2-B\frac{k^2}{a^2}-\frac{1}{\Lambda_c^2}\(\omega^2-\frac{k^2}{a^2}\)^2=0\,,
\ea
where we work with modes well above the background and hence the mass, so the mass term is safely ignored in the previous expression.
The previous equation can be solved as
\ba
\label{eq:omegaGhost}
\omega^2(k)=c_s^2 \frac{k^2}{a^2}+\frac{(A-B)^2}{A^3}\frac{k^4}{a^4\Lambda_c^2} +\mathcal{O}\(k^6/\Lambda_c^4\)\,.
\ea
In general the first term dominates in the naive region of validity of the EFT $k/a \ll \Lambda_c$. However, for the bouncing solutions $c_s^2$ could become very small, potentially passing through zero and becoming negative, and the second term then regulates any instability that may occur, while the higher order corrections $\mathcal{O}(k^6/\Lambda_c^4)$ remain small in comparison.
Interestingly in our explicit solutions, we will find that the inclusion of the operator $(\Box \Phi)^2/\Lambda_c^2$ also modifies the background solution in such a way that $c_s^2$ remains positive throughout the bounce. Again this is achieved while terms like $(\Phi \Box^{n+2} \Phi)/\Lambda_c^{2+2n}$ remain negligible. \\

In addition to the dispersion relation \eqref{eq:omegaGhost}, there is a second solution which would be the Ostrogradsky ghost mode
\ba
\omega_{\rm ghost}^2 \sim A \Lambda_c^2 \(1+\mathcal{O}\(\frac{k^2}{\Lambda_c^2}\)\)\,.
\ea
These non-unitary degrees of freedom must be excluded from the EFT, and since they enter at an energy scale $\sqrt{A}\Lambda_c$, this sets the maximal value of the energy cutoff of the theory on this background,
\ba
\Lambda_{\rm cutoff} \le  \Lambda_c'=\sqrt{A}\Lambda_c\,.
\ea
In addition to $\Lambda_{\rm cutoff}$, it will be useful in what follows to introduce the scale $\mu_c$ which is the energy scale at which the dispersion relation transitions from the relativistic form $\omega = c_s k$, to the non-relativistic form dominated by the operator $(\Box \Phi)^2/\Lambda_c^2$:
\ba
\label{eq:mucFull}
\mu_c=c_s^2 \frac{A^{3/2}}{(A-B)} \Lambda_c\,.
\ea
Due to the background redressing,  $\mu_c$ may differ significantly from the scale $\Lambda_c$ and from the cutoff of the theory. For instance if we were in a situation where $B \ll 1\ll A$ then $\mu_c\sim B/\sqrt{A} \Lambda_c\ll \Lambda_c \ll \Lambda_{c}'$.    In practice, the situations where the irrelevant operator has an effect on the dispersion relation arise because the speed of sound $c_s^2=B/A$ is small, which implies that $B\ll A$. In what follows we shall slightly simplify the notation by making that assumption and hence write,
\ba
\label{eq:mucSimp}
\mu_c \sim c_s^2 \sqrt{A} \Lambda_c = \frac{B}{\sqrt{A}} \Lambda_c\,.
\ea
From the quadratic action \eqref{eq:S2} (or the dispersion relation \eqref{eq:omegaGhost}), it is clear that when considering modes $k/a \gg \mu_c/c_s$ -- or equivalently when looking at energy scales $\omega \gg \mu_c$ -- the irrelevant operator takes over from the standard gradient term. There are now therefore two different regimes to consider when checking perturbative unitarity of the theory.

\subsection{Intermediate energy modes with $E_{\rm back}^2 \ll s \ll \mu^2_c $}
\label{sec:intermediate}
For modes with energy well below the scale $\mu_c$ (but yet well above any of the background scales) the high energy effects are irrelevant and we can perform the analysis in the same way as we did in the previous section in the absence of the irrelevant operator $(\Box \Phi)^2/\Lambda_c^2$. We can therefore infer that:
 \begin{itemize}
 \item If any of the energy scales $\mu_{NML}$ derived in \eqref{eq:muNML} happened to be smaller than the energy scale $\mu_c$, then the smallest of those scales would be the maximal value of the strong coupling scale and perturbative unitarity would break down at that scale.
\item If on the other hand {\it all} the energy scales $\mu_{NML}$ derived in \eqref{eq:muNML} are larger than $\mu_c$ then higher energy effects (which are still below the cutoff) ought to be taken into account to properly diagnose the scale of perturbative unitarity breaking and the scales computed in  \eqref{eq:muNML} should then be ignored.
\end{itemize}

\subsection{High energy modes with $\mu^2_c \ll s \ll \Lambda^2_{\rm cutoff} $}
\label{sec:highenergymodes}

For the high energy modes, $\mu^2_c \ll s \ll \Lambda^2_{\rm cutoff} $ as we have seen already at the quadratic level, the irrelevant operator $k^4 \varphi^2/\Lambda_c^2$ takes over from the standard gradient term.
At those energy scales, the second contribution in \eqref{eq:omegaGhost} dominates the dispersion relation which hence becomes
\ba
\omega\sim \frac{1}{A^{1/2}\Lambda_c}\frac{k^2}{a^2}\sim \frac{c_s^2}{\mu_c }\frac{k^2}{a^2}\,.
\ea

\paragraph{Scaling:}

The scaling performed in section \eqref{sec:scalarFLRW} should hence be reinvestigated. Considering the operator $(\p^2_i \varphi)^2/\Lambda_c^2$ rather than the subdominant gradient term $B (\p_i \varphi)^2$, the appropriate rescaling would then correspond to substituting $B$ by $(k/a \Lambda_c)^2\sim(A s/\Lambda_c^2)^{1/2}$. This is a slight abuse of notation, however when working in the WKB regime, such a rescaling can be justified.
This means that an operator of the form \eqref{eq:OperatorsNML} (with $N+2M+4L>4$) preserves perturbative unitarity so long as the energy $\sqrt{s}$ satisfies
\ba
\label{eq:highenergyrel}
\mu_c \lesssim \sqrt{s}\ll A^{n_A} \(\frac{A s}{\Lambda_c^2}\)^{n_B/2} \Lambda_{NML}\,,
\ea
where the powers $n_A$ and $n_B$ are given in \eqref{eq:nAnB}.
If $n_B < 1$, \ie for $N+5M+6L>10$ this implies unitarity is broken above the energy
\ba
\label{eq:strongMuhigh}
\sqrt{s} \sim A^{\frac{N+M-2L+6}{2(N+5M+6L-10)}}\(\frac{\Lambda_{NML}}{\Lambda_c}\)^{\frac{4(N+2M+4L-4)}{N+5M+6L-10}} \Lambda_c \quad{\rm for }\quad N+5M+6L>10\,.
\ea
On the other hand for $n_B\ge 1$, the bound \eqref{eq:highenergyrel} is always satisfied so long as it is satisfied at $s=\mu_c^2$, \ie so long as
\ba
\label{eq:LambdaNMLbound}
\Lambda_{NML}\gg A^{-\frac{2N+3M+6L-4}{2(N+2M+4L-4)}} B^{\frac{7N+11M+11L-22}{4(N+2M+4L-4)}} \Lambda_c\quad{\rm for }\quad N+5M+6L\le10 \,,
\ea
(and for $N+2M+4L>4$). For the relevant and marginal operators $\varphi^3$ and $\varphi^4$, this distinction between intermediate and low energy modes is unimportant and we can follow the discussion of Eqns.~(\ref{eq:mu300}-\ref{eq:mu400}).
\vspace{0.5cm}

\paragraph{Scattering Amplitude:} The previous argument required a scaling which itself depended on momenta. For a more rigorous treatment, one can instead first directly compute the bounds imposed by the optical theorem in the presence of the irrelevant operator and then estimate the $2n$  point functions without the need of ever performing the rescaling of the coordinates mentioned previously. Just as in section~\ref{sec:Optical1}, both methods give the same result, but we sketch the direct bounds from the optical theorem (without the need of any rescaling) to solidify the argument and the results.\\

When including the irrelevant operator, first notice that the correct normalization of the field in \eqref{eq:modesphi}
is still given by $\mathcal{N}=\frac{\p F}{\p \omega}$ as in the first equality of Eq.~\eqref{eq:Norm1} but where the function $F$ is now given by \eqref{eq:DispF2} and where the dispersion relation $\omega=\frac{1}{A^{1/2} \Lambda_c }\frac{k^2}{a^2}$, hence leading to
\ba
\mathcal{N}=\frac{\p F}{\p \omega} = \frac{2}{A^{5/2} \Lambda_c} \frac{k^2}{a^2}\,.
\ea
This change of normalization $\mathcal{N}$ affects the momenta integrals in \eqref{eqn:A2n1} which in turn affects the bounds provided by the optical theorem.
Taking these effects into account, the perturbative unitarity bound from the $2n$-point functions \eqref{eqn:A2nboundlow1} now becomes
\begin{equation}
 | \A_{2n} |  \lesssim  A^{(n+3)/4} \(\frac{\sqrt{s}}{\Lambda_c}\)^{\frac{3(n-1)}{2}} s^{2-n} \,,
\label{eqn:A2nboundlow2}
\end{equation}
again up to numerical factors that have been ignored.\\

Now if we consider the effect of an operator given in \eqref{eq:OperatorsNML}, with $N+M+2L=2n$ and $N+2M+4L>4$, it will lead to a contribution to the $2n$-point function going as
\ba
\label{eq:ANMLg}
\A_{NML}\sim \frac{k^{2L}\omega^M}{\Lambda_{NML}^{N+2M+4L-4}}\sim A^{L/2}\frac{\Lambda_c^L s^{(M+L)/2}}{\Lambda_{NML}^{M+2L+2n-4}}\,.
\ea
Of course this is only one contribution to that amplitude, and this does not account for the very special case where this contribution happens to be precisely canceled that from another operator. This would mean that these two (or more) operators are removable by field redefinition. This requires a very precise tuning between these operators and simply means that the original formulation of the theory was unnecessarily complicated. We deal with this here by defining the relevant scale $\Lambda_{NML}$ \emph{after} all the appropriate and relevant field redefinitions have been performed. See section~\ref{sub:fieldRedef} for more discussion on this point. We have also ignored any combinatory factors that ought to be included when computing the amplitude, just like we have ignored for now any factorial that enters in the definition of the operator \eqref{eq:OperatorsNML} but those can easily be accounted for once we have diagnosed the potentially dangerous operators and amplitudes.  \\

Then from Eq.~\eqref{eqn:A2nboundlow2}, the contribution to amplitude \eqref{eq:ANMLg} would diagnose a violation of tree-level unitarity at the energy scale
\ba
\sqrt{s}\sim A^{\frac{3-2L+2n}{2(-5+2L+2M+n)}}\(\frac{\Lambda_{NML}}{\Lambda_c}\)^{\frac{2(-4+2L+M+2n)}{-5+2L+2M+n}}\Lambda_c\,,
\ea
for $N+2M+2L>5$ (or $N+5M+6L>10$),
which is precisely the scale inferred from the simple scaling argument in \eqref{eq:strongMuhigh} when using  $N+M+2L=2n$. For operators with $N+5M+6L\le10$ on the other hand, they respect unitarity so long as they satisfy the same bound as that derived in \eqref{eq:LambdaNMLbound}.  \\

\paragraph{Strong Coupling Scale:} To summarize, we have computed in two different and yet complementary ways the strong coupling scale associated with a class of scalar EFTs on a cosmological background. We consider any scalar field EFT that takes the following form on a cosmological background,
\ba
S=\int \d^4 x \, a^3 \Bigg(\frac A2\, \dot \varphi^2 - \frac{B}{2}\frac{(\p_i \varphi)^2}{a^2} -\frac 12 m^2 \varphi^2
+\hspace{-3mm}\sum_{\substack{N,M,L\ge 0 \\ N+M+2L\ge 3 }}\frac{\varphi^N \dot \varphi^M (\p_i \varphi)^{2L}}{\Lambda_{NML}^{N+2M+4L-4} a^{2L} }  +\frac{\left(  \Box \varphi  \right)^2}{2\Lambda_c^2}
\Bigg)\,,\qquad
\ea
where $A,B,m$ and the scales $\Lambda_{NML}$ are all functions of time and depend on the precise theory one is dealing with and on the particular cosmological background considered. The scales $\Lambda_{NML}$ are the relevant scales of the theory on the cosmological background after all the appropriate and relevant field redefinitions have been performed and typically $\Lambda_{NML}\ll A \Lambda_c$. We denote by $E_{\rm back}$ the scale of the background. \\

Then for that theory,  (modulo the subtleties related to field redefinitions and redundant operators discussed in section~\ref{sub:fieldRedef}) perturbative unitarity requires the following conditions on the scale of the different irrelevant operators:
\begin{enumerate}
\item {\bf For any  irrelevant operator} ($N+2M+4L>4$),
\ba
\label{eq:req2}
A^{- \frac{N+M+6L-6}{4(N+2M+4L-4)}}B^{\frac{3N+3M+10L-6}{4(N+2M+4L-4)}} \Lambda_{NML} \gg   \mu_c \sim \frac{B}{\sqrt{A}} \Lambda_c
\ea
\item {\bf For all irrelevant operators} with $N+5M+6L>10$,
\ba
\label{eq:req3}
 A^{\frac{N+M-2L+6}{2(N+5M+6L-10)}}\(\frac{\Lambda_{NML}}{\Lambda_c}\)^{\frac{4(N+2M+4L-4)}{N+5M+6L-10}} \Lambda_c \gg E_{\rm back}
 \ea
 \item {\bf For all the irrelevant operators} with $N+2M+4L-4>0$ and $N+5M+6L\le 10$,
\ba
\label{eq:req4}
A^{\frac{2N+3M+6L-4}{2(N+2M+4L-4)}} B^{-\frac{7N+11M+11L-22}{4(N+2M+4L-4)}} \Lambda_{NML} \gg   \Lambda_c \,.
\ea
\end{enumerate}
The strong coupling scale is designated by
\ba
\label{eq:Lambdas}
\Lambda_s={\rm Min}_{N,M,L}\(A^{\frac{N+M-2L+6}{2(N+5M+6L-10)}}\(\frac{\Lambda_{NML}}{\Lambda_c}\)^{\frac{4(N+2M+4L-4)}{N+5M+6L-10}}\) \Lambda_c\,,
\ea
where the minimum is taken over all the $N,M,L$ which satisfy $N+5M+6L>10$ and
where we have implicitly assumed  that $E_{\rm back}\gtrsim \mu_c\sim B/\sqrt{A} \Lambda_c$ and the second condition \eqref{eq:req2} is satisfied. If $E_{\rm back}\lesssim \mu_c$, the last two requirements would be irrelevant and the right hand side of \eqref{eq:req2} would simply be $E_{\rm back}$ instead of $\mu_c$ and the strong coupling scale of interest would be instead given by the left hand side of \eqref{eq:req2}. Once again, these conditions should be taken as a diagnosis. If any of those were broken one could go back to computing the explicit contributions to the scattering amplitudes which would account for the proper numerical factors and is independent of field redefinitions.  \\

When it comes to the relevant and marginal operators, since those operators are renormalizable, one can in principle  deal with those in the strong coupling regime. Remaining in the perturbative regime requires
\ba
\label{eq:req1}
\mu_{300}=  \frac{\Lambda_{300}}{(AB)^{3/4}} \lesssim  E_{\rm back}\quad {\rm and } \quad \mu_{400}= \frac{\lambda}{(A B^3)^{1/2}} \lesssim \mathcal{O}\(16 \pi\)\,,
\ea
(where $\lambda$ is the coupling constant that enters for the $\varphi^4$ operator); however we emphasize that breaking this bound \eqref{eq:req1} does not directly imply breaking of unitarity, but rather that loops from these operators ought to be considered. \\

In the rest of this manuscript, we shall now use these bounds to infer if and how a NEC violation can occur in $P(\Phi,X)$ theories of gravity and whether a bounce is possible without violation of unitarity.

\section{Violating the NEC in $P(\Phi,X)$ theory}
\label{sec:PX}

We now focus the discussion of perturbative unitarity on $P(\Phi,X)$ theories near a NEC violating region. In that case the scalar field Lagrangian in \eqref{eq:Soriginal} takes the form
\ba
S[g\mn, \Phi]=\int \d^4 x \sqrt{-g}\(\frac{\mpl^2}{2}R +P(\Phi,X)   \) \,,
\ea
with
\ba
 X=-\frac12 g^{\mu\nu} \p_\mu \Phi \p_\nu \Phi\,,
\ea
and as before, we will be interested in the effective scalar field theory for $\varphi$ on the FLRW cosmological background where we have set $\Phi=\phi+\varphi(t,x^i)$ and the background satisfies the appropriate equations of motion. In particular the background energy density is given by
\ba
\rho=- \bar P + \bar P_{,X} \dot{\phi}^2\,,
\ea
where `bar' quantities are related to the background, \ie $\bar X= \frac12 \dot{\phi}^2$ and $\bar P=P(\phi, \bar X)$. \\

The kinetic coefficients and the mass that determine the quadratic action for $\varphi$ are given by
\ba
A= 2 \bar X \bar P_{, X X}+ \bar P_{, X}\,, \qquad B= \bar P_{, X} \qquad {\rm and }\qquad m^2=\p_t[\bar P_{,\Phi X} \dot{ \phi}]-\bar P_{,\Phi \Phi}\,,
\ea
and stability of the theory requires these three functions to be positive.
Note that when accounting for the kinetic factors $A$ and $B$, the effective mass scale `perceived' by the properly normalized scalar field is  given by
\ba
\label{eq:meff}
m_{\rm eff}^2= A^{-1} m^2 = A^{-1} \(\p_t[\bar P_{,\Phi X} \dot{ \phi}]-\bar P_{,\Phi \Phi}\)\,.
\ea
The operators $S_{NML}$ \eqref{eq:OperatorsNML} of the effective theory on the cosmological background can be derived for any given $P(\Phi,X)$ model, and the scales $\Lambda_{NML}$ are given by
\ba
\label{eq:LambdaNML PX}
\Lambda^{-(N+2M+4L-4)}_{NML} = \sum_{j=0}^{M/2}\frac{(-1)^L \dot{\phi}^{M-2j} }{2^{j+L}N! L! j! (M-2j)!} \p_{\Phi}^N
\p_X^{M+L-j} \bar P\,.
\ea
Without having more insight on the precise form of the function $P$ and whether or not it truncates at any order in $\Phi$ or $X$, there is little more one can say about those scales. In what follows we will hence take into account the fact that the function $P$ allows for a NEC violation, and even consider the case where $P$ is such that a cosmological bounce occurs. We will first focus on the case without any higher order effects before including their effects.

\subsection{Ignoring high energy effects}

When considering a pure $P(\Phi,X)$ model, without the inclusion of higher energy effects as mimicked by the operator $(\Box \Phi)/\Lambda_c^2$ in section~\ref{sec:HighEnergy}, the background equations of motion are simply
\ba
\label{eq:FriedPX}
&& 3 \mpl^2 H^2 = \rho = - \bar P + 2\bar P_{,X} \bar X \, , \\
&& \mpl^2 \dot H = -\frac 12 (p+\rho)= - \bar X \bar P_{,X} \,.
\label{eq:raychaudhuri}
\ea
It is well-known that in a $P(\Phi,X)$ theory, at the onset of a NEC violation\footnote{From the Raychaudhuri Eq.~\eqref{eq:raychaudhuri} we could try to onset the NEC by setting $\dot{\phi}=0$ rather than $\bar P_{, X} = B = 0$. However setting  $\dot{\phi}=0$  would also imply $\ddot H=0$ at that time, which would mean that we are not within the NEC violating region unless one also had $\dddot H=0$ at that time, which itself also implies $\bar P_{, X}=0$ (or $\ddot{\phi}=0$, in which case the same story continues). So to summarize, starting outside the NEC region $\dot H<0$, one can never enter within the NEC region $\dot H>0$ without passing through a point for which $\bar P_{, X} =0$ in a pure $P(\Phi,X)$ theory. We can therefore assume that $\dot{\phi}\ne 0$ at the onset of the NEC violation without loss of generality.}  $\bar P_{, X} = B = 0$. Even more worrisome than the standard stability issues caused by the gradient terms vanishing or becoming negative, we can directly see that the onset of the NEC violation in any `pure' $P(\Phi,X)$ theory is inexorably linked with a violation of unitarity which makes it impossible to trust the fact that the NEC violation actually occurred in the first place. \\

Indeed, if $B\to 0$ the strong coupling scale derived in \eqref{eq:muNML} (or in \eqref{eq:Lambdas}) vanishes (hence signaling violation of perturbative unitarity at an arbitrarily small scale) unless {\it all} the operators with $3N+3M+10L>6$ vanished (or could all have been simultaneously removed by a field redefinition) which means there can be no interactions at all, which of course means there could not have been any operator to set a NEC violation in the first place. So without including higher energy effects, even putting aside any stability issue that may occur at the classical level, there can be no NEC violation within the regime of validity of a pure $P(\Phi, X)$ model.  In what follows we shall therefore include the high energy effects that are naturally expected to be present in any effective field theory. Notice that we still work within the low-energy effective field theory and therefore should not be sensitive about the exact details of the UV physics nor its exact realization, but their effect plays the role of a regulator for the low-energy effective theory which are crucial for the onset of a NEC violation in $P(\Phi,X)$ types of theories.

\subsection{Including high energy effects}
\label{sec:highEnergyPX}
When including the operator $(\Box \Phi)^2/\Lambda_c^2$, \ie when considering the following action,
\ba
S[g\mn, \Phi]=\int \d^4 x \sqrt{-g}\(\frac{\mpl^2}{2}R +P(\Phi,X)  +\frac{1}{2\Lambda_c^2} (\Box \Phi)^2 \) \,,
\ea
the background equations of motion are slightly modified to
\ba
\label{eq:FriedPXLambdac}
&& 3 \mpl^2 H^2 = - \bar P +2 \bar P_{,X} \bar X+\frac{1}{2\Lambda_c^2}\left[-\ddot \phi^2+2\dot \phi \dddot \phi+6 \dot H\dot \phi-9H^2 \dot \phi^2 \right] \, , \\
&& \mpl^2 \dot H =  - \bar X \bar P_{,X} +\frac{1}{\Lambda_c^2} \left[\dot \phi \dddot \phi + 3\dot H \dot \phi^2+3 H \dot \phi \ddot \phi  \right]\,.
\label{eq:raychaudhuriLambdac}
\ea
 The effect of $\Lambda_c$ on the Friedman equation is irrelevant when working within the regime of validity of the effective field theory, but it allows for a violation of the NEC at finite positive $B$,
\ba
B=\bar P_{,X}=\frac{2}{\Lambda_c^2 \dot{\phi}}\(\dddot{{\phi}}+3 H \ddot{\phi}\)>0 \qquad {\rm when }\qquad \dot H=0\,.
\ea
As mentioned in \eqref{eq:Ebackground2}, the scale of the background, $E_{\rm back}$, is given by
\ba
\label{eq:Ebackground}
E_{\rm back} = {\rm Max}\(H, \sqrt{\dot H}, m,  \dot \phi/ \phi, \ddot \phi / \dot \phi, \cdots \)\gtrsim \sqrt{B} \Lambda_c \,,
\ea
where the last inequality is valid when $\dot H\sim0$. Preserving perturbative unitarity requires the background scale to be smaller that the strong coupling scale derived in \eqref{eq:Lambdas}
\ba
E_{\rm back} \ll \Lambda_s
\ea
which requires
\ba
\label{eq:UnitarityReq}
\sqrt{B} \Lambda_c \ll \Lambda_s\,.
\ea
Some of the literature does also require the background energy density $\rho$ to be much smaller than the strong coupling scale of the perturbed effective field theory $\rho \ll \Lambda_s^4$. This is certainly a safe requirement to impose as one would not {\it expect} the EFT to be stable under quantum corrections otherwise. From a pure unitarity-preserving perspective however, we are not forced to impose that condition.  \\

Since we can assume $\dot{\phi} \ne 0$, and since the derivatives of the background should be small compared with the scale of high energy physics $\Lambda_c$, we can immediately infer the hierarchy
\ba
B \ll A \quad {\rm and }\quad B\ll 1 \qquad {\rm when }\qquad \dot H=0\,,
\ea
which implies that the speed of sound ought to be small in this model for a NEC violation to occur. A small sound speed may not necessarily be problematic at the classical level, especially not when the high energy effects regulate any gradient instability that could occur, but as we shall see it greatly affects unitarity.

To establish whether or not perturbative unitarity can be preserved in the vicinity of a NEC violating region we start by looking at the effect of the intermediate modes as described in section~\ref{sec:intermediate}. We therefore consider modes with energy scale $E_{\rm back}^2\ll s \ll \mu_c^2$. In that regime, if any of the scales derived in \eqref{eq:muNML} happened to be smaller than $\mu_c\sim B A^{-1/2} \Lambda_c \ll \Lambda_c$, with now the $\Lambda_{NML}$ given by \eqref{eq:LambdaNML PX} (and keeping in mind that $B\ll A \sim 2 \bar X \bar P_{,XX}$), then that scale would set the strong coupling scale. \\

\paragraph{Breaking the shift symmetry:}

Let us start by considering a $P(\Phi,X)$ theory which does not necessarily preserve the shift symmetry (\ie has explicit $\Phi$ dependence).  We also start by  making the very natural assumption   that the fundamental theory does not carry any hierarchy of scales by which we mean that  one can formulate the function $P(\Phi,X)$ in terms of just one scale $\Lambda (\ll \Lambda_c \ll \mpl )$ and order one dimensionless coefficients $c_{\ell,m}$,
\ba
\label{eq:OneScaleP}
P(\Phi,X)=\Lambda^4\sum_{\ell, n} \frac{c_{\ell,n}}{\Lambda^{\ell+4n}}\Phi^\ell X^n\,.
\ea
We would then have $\p^4_\Phi \bar {P} \sim 1$ (or $\p^4_\Phi \bar {P} \gg 1$ if the background involved $\phi\gg \Lambda$ or $\bar X\gg \Lambda^4$, but we would not be able to have $\p^4_\Phi \bar {P} \ll 1$  unless a very specifically tuned cancellation occurred precisely at the onset of the NEC violation). Similarly we would never expect to have $\Lambda^{N+4M-4}\p_\Phi^N\p_X^M \bar P \ll 1$ unless a very particular tuning was set to occur precisely at the onset of the NEC violation, or unless the shift symmetry or another precise type of symmetry was present. So in a typical theory \eqref{eq:OneScaleP} with no hierarchy of scales, we expect $\Lambda_{NML}\sim \Lambda$ for all $N+2 M+4 L\ne 4$.\\

With this assumption in mind, we can consider the quintic operator
\ba
\L_{500}= \frac{\varphi^5}{\Lambda_{500}}\qquad{\rm with}\qquad \Lambda_{500}=5! \( \p_\Phi^5 \bar P\)^{-1}\sim 5! \Lambda\,.
\ea
As we have seen, to avoid any breaking of unitarity this operator should satisfy the requirements \eqref{eq:req2} and \eqref{eq:req4}, leading to
\ba
 \Lambda \gtrsim (A^3 B^5)^{-1/4} \Lambda_c \,,
\ea
which further requires $A \gtrsim B^{-5/3} (\Lambda_c/\Lambda)^{4/3} \gg 1$, which is only possible if some of the coefficients present in  \eqref{eq:OneScaleP} are much larger than unity.
Once the door is opened for such a special tuning (\ie when some coefficients $c_{\ell, m}$ in \eqref{eq:OneScaleP} are allowed to be parametrically much larger than others), any vertex can in principle dominate the scattering amplitudes and lead to much stronger bounds than would be inferred from the other vertices.

\paragraph{Preserving the shift symmetry: } An obvious way to evade the previous argument is to keep the shift symmetry and hence avoid any operator that depends explicitly on $\varphi$. We can start by assuming as we did earlier  that the covariant theory contains no large hierarchies,
\ba
\label{eq:OneScalePX}
P(X)=\Lambda^4\sum_{n\ge 1} \frac{c_{n}}{\Lambda^{4n}} X^n\,,
\ea
with all the $c_n$ of order 1, with the possibility that some of them may vanish and that the sum may truncate at order $\bar N$. In that case one should have  $\Lambda^{4(n-1)}\p_X^n \bar P \sim  1$ (or $\gg 1$)  for any $2\le n\le \bar N$ unless there is particular artificial tuning (or hierarchy of scales). In this case the most stringent bounds come from the cubic and quartic operators $\dot \varphi (\p_i \varphi)^2$ and $(\p_i \varphi)^4$ which impose the requirement \eqref{eq:req2} $\Lambda\gtrsim c_s^{1/4} \Lambda_c$.
Since $c_s\ll 1$ at the onset of the NEC violation in these types of theories, this requirement is not a priori unreasonable, but it does constrain the theory.  Once this constraint is satisfied, none of the other operators  of the theory would break perturbative unitarity so long as no artificially large coefficient is included in $P(X)$. Expressed as a constraint on $P(X)$ we see that perturbative unitarity requires
\ba
1\ll \Lambda_c^4 \bar P_{,XX} \ll  c_s^{-1}\,,
\label{eq:PXX}
\ea
(where the lower bound is coming from the requirement that $\Lambda \ll \Lambda_c$). The pure ghost-condensate \cite{ArkaniHamed:2003uy,Creminelli:2006xe,Buchbinder:2007ad,Koehn:2015vvy} lies in that category of models and is discussed in more detail in appendix~\ref{App:GC}. For that model we see that the unitarity bounds put severe constraints on the high energy operators which end up needing to break the shift symmetry.  \\

The previous unitarity bound was derived by estimating the contribution of the cubic and quartic vertices to the $2-2$ scattering amplitudes. Having identified the potentially most dangerous vertices, we can go ahead and compute their actual contributions to the tree-level scattering amplitudes to ensure that no `accidental cancellations' occur. The direct calculation of the $2-2$ tree scattering amplitude taking into account both the cubic and quartic vertices is in complete agreement with the estimations and provides the following upper bound,  $ \Lambda_c^4 \bar P_{,XX} \ll 192 \pi^2/(103  c_s)$.  \\

\paragraph{Summary:} To summarize, a NEC violation in shift-symmetric $P(X)$ theory is in principle possible so long as higher energy effects enter at a sufficiently low energy scale to regulate the scattering amplitudes, but still at sufficiently high energies so as not to entirely spoil the low-energy EFT. With these conditions in mind we obtain a limited, but not necessarily empty, window of possibility. \\

In practice however, when it comes to obtaining explicit bouncing solutions, breaking the shift symmetry may make the bounce `easier' to model. In the absence of a protecting shift symmetry, the unitarity bounds are tighter and require an additional level of tuning of the model. These additional tunings imply that the $2-2$ scattering is no longer necessarily the dominant scattering amplitude and all processes should be examined with care to determine whether unitarity is preserved. One should also ensure that the effective mass and couplings of the marginal and relevant operators are sufficiently small.
 This can be done explicitly, and to illustrate the process we now provide an explicit model which allows for a stable cosmological bounce that preserves unitarity at the price of introducing an unnaturally small parameter.

 \subsection{Explicit Model}

\label{Explicitmodel}

We now present an explicit model of the form
\ba
S[g\mn, \Phi]=\int \d^4 x \sqrt{-g}\(\frac{\mpl^2}{2}R +P(\Phi,X)  +\frac{1}{2\Lambda_c^2} (\Box \Phi)^2 \) \,,
\ea
with
\begin{equation}
P ( \Phi, X ) = - \Lambda^4 V (\Phi) + p (\Phi) X + \frac{q ( \Phi)}{\Lambda^4} X^2\,,
\label{eqn:egPX}
\end{equation}
so the background equations of motion are given by
\ba
\label{eq:FriedPXLambdac2}
&& 3 \mpl^2 H^2 = \Lambda^4 V+ p \bar X +\frac{3}{\Lambda^4}q \bar X^2+\frac{1}{2\Lambda_c^2}\left[-\ddot \phi^2+2\dot \phi \dddot \phi+6 \dot H\dot \phi-9H^2 \dot \phi^2 \right] \, , \\
&& \mpl^2 \dot H = -p \bar X-2\frac{q}{\Lambda^4}\bar X^2 +\frac{1}{\Lambda_c^2} \left[\dot \phi \dddot \phi + 3\dot H \dot \phi^2+3 H \dot \phi \ddot \phi  \right]\,.
\label{eq:raychaudhuriLambdac2}
\ea
We can then explicitly check that the following profile
\begin{equation}
\dot \phi = \Lambda \phi , \;\;\;\; H =\frac{\Lambda^3}{\mpl^2} h (\phi )\,.
\end{equation}
is an exact solution of the background equations of motion if we choose the following potential and functions of $\Phi$
\begin{align}
V ( \Phi ) &= -\frac 14 q(\Phi) \frac{\Phi^4}{\Lambda^4} +  \left( 3\frac{\Lambda^2}{\mpl^2} h^2 (\Phi ) + \Phi h' (\Phi) \right) - \frac{\Phi^2}{2 \Lambda_c^2} \left( 1 + 3\frac{\Lambda^2}{\mpl^2} h (\Phi ) \right)^2 \, ,  \\
p (\Phi ) &= -  q(\Phi)\frac{\Phi^2}{\Lambda^2} -2 \frac{ \Lambda^2}{\Phi} h' (\Phi) + \frac{2 \Lambda^2}{\Lambda_c^2} \left[ 1 + 3 \frac{\Lambda^2}{\mpl^2} h (\Phi) + 3 \frac{\Lambda^2}{\mpl^2} \Phi h' (\Phi) \right] \, ,
\label{eqn:PX_direct}
\end{align}
and for any pair of free functions $q (\Phi)$ and $h(\Phi)$. This freedom ensures that we can choose a profile that undergoes a cosmological bounce while preserving unitarity. \\

Focusing on the decoupling limit about the FLRW background as discussed in section~\ref{sec:dL},
the coefficients of the kinetic matrix are then given by
\ba
A&=&2q(\phi) \frac{\phi^2}{\Lambda^2}-2\frac{\Lambda^2}{\phi}h'(\phi)+2\frac{\Lambda^2}{\Lambda_c^2}
\left[1+3\frac{\Lambda^2}{\mpl^2}\(h(\phi)+\phi h'(\phi)\)\right] \, ,\\
B&=&-2\frac{\Lambda^2 h'(\phi)}{\phi}+2\frac{\Lambda^2}{\Lambda_c^2}
\left[1+3\frac{\Lambda^2}{\mpl^2}\(h(\phi)+\phi h'(\phi)\)\right]\,,
\ea
and so the sound speed is given by
\ba
c_s^2=\frac{\phi h'(\phi)-\frac{\phi^2}{\Lambda_c^2}}{\phi h'(\phi)- q \frac{\phi^4}{\Lambda^4}- \frac{\phi^2}{\Lambda_c^2}} +\mathcal{O}\(\frac{\Lambda}{\mpl}\)\,.
\ea

\begin{figure}[h]\vspace{-2cm}
\centering
\includegraphics[width=0.75\textwidth]{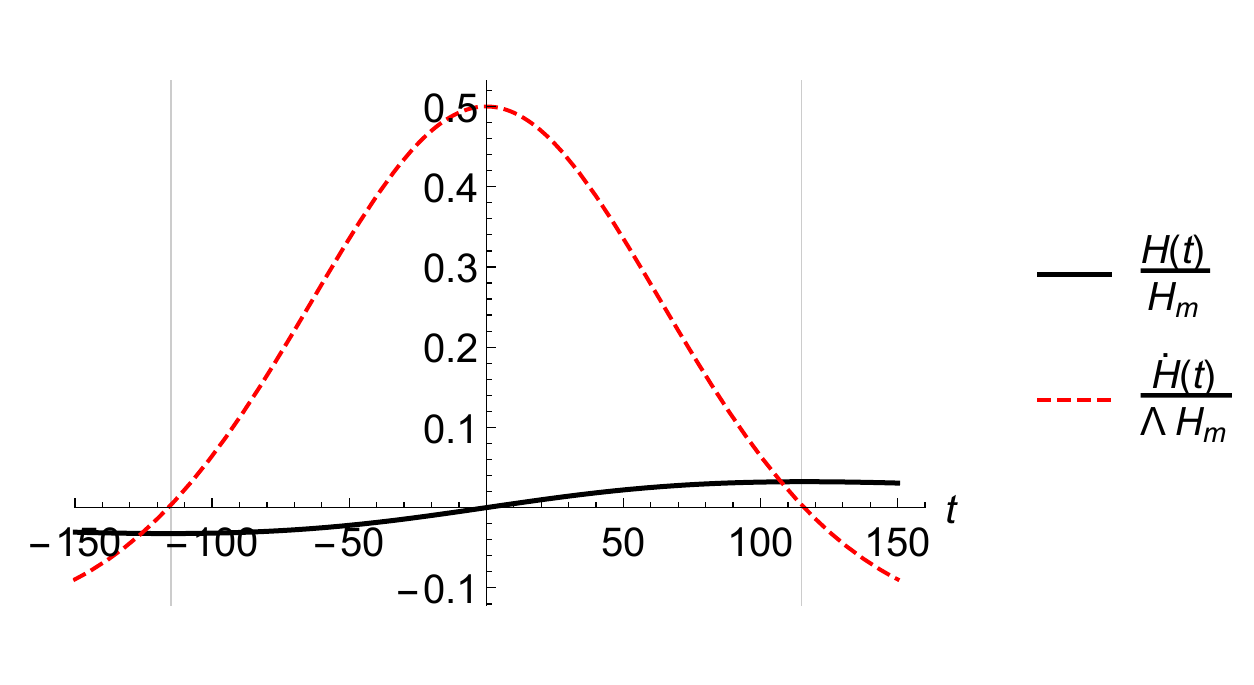}
\includegraphics[width=0.75\textwidth]{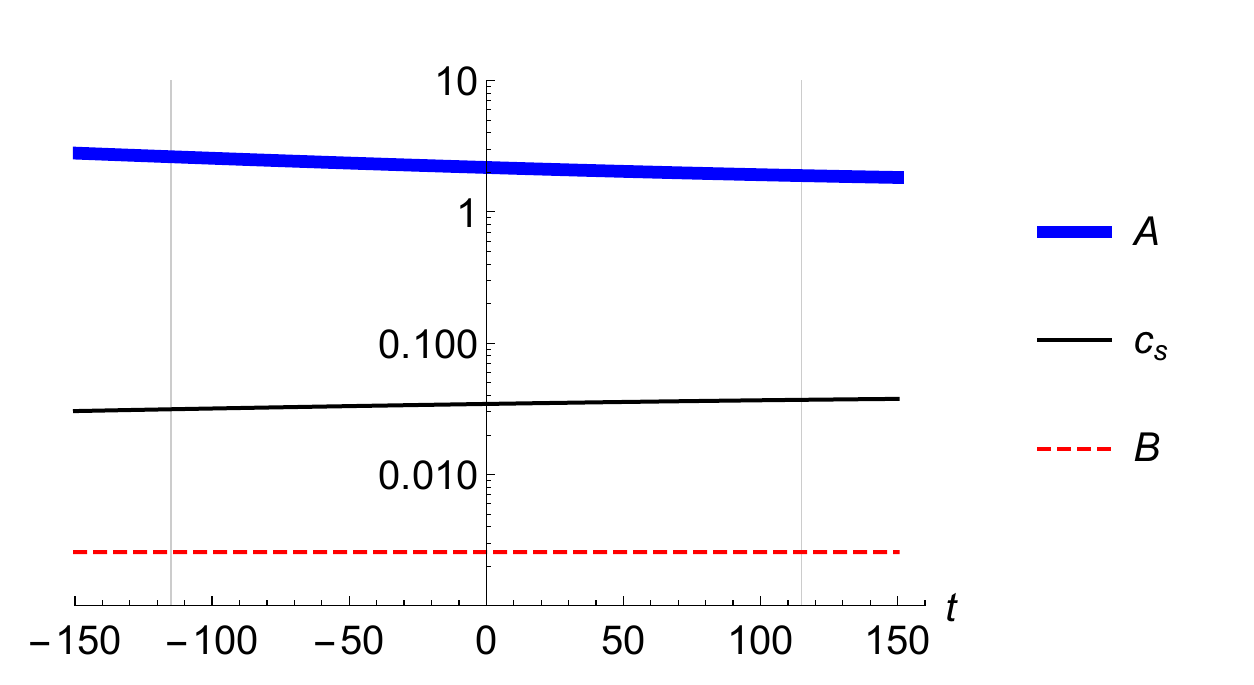}
\includegraphics[width=0.75\textwidth]{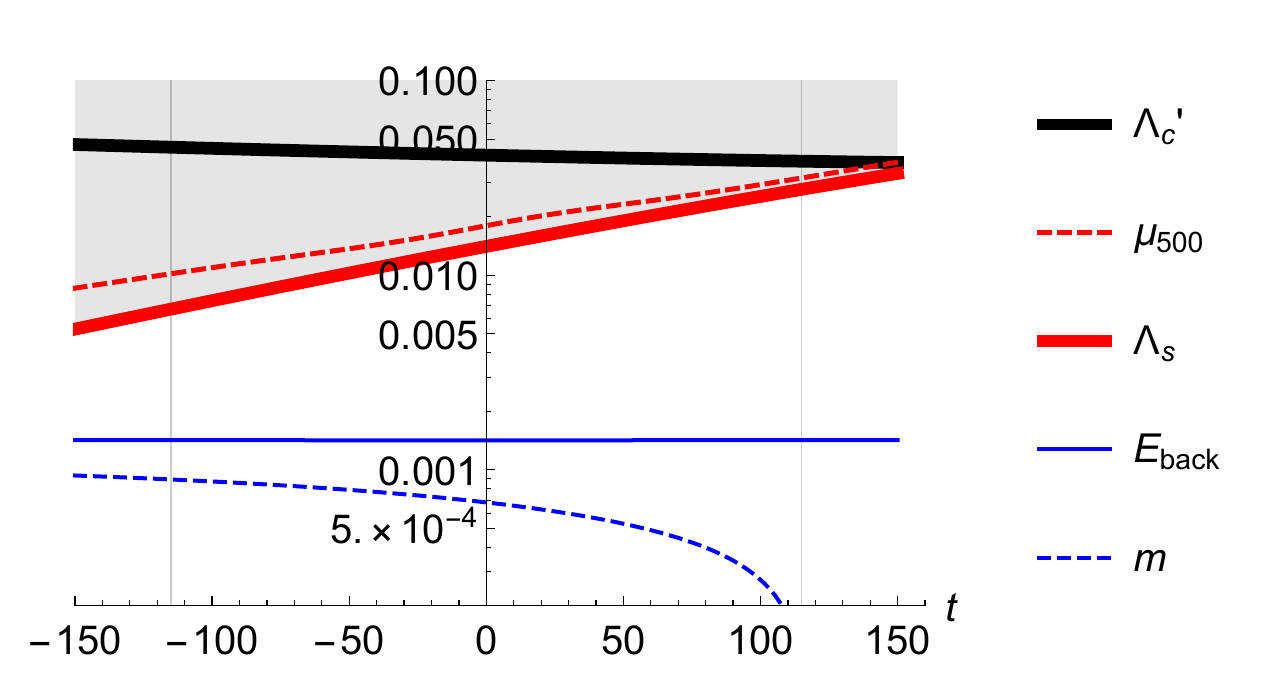}
\caption{The behaviour of the specific example provided in \eqref{eqn:PXsoln_direct}. The time is given in Planck scale units. The Hubble parameter is measured in terms of the scale $H_m=\Lambda^3/\mpl^2$.  The sound speed is manifestly positive throughout the NEC violating region. On the lower plot, the relevant scales of the system are represented relative to $\mpl$.
$\Lambda_c'$ represents the scale at which the higher energy effects enter, $\rho$ is the energy density of the background, $m$ is the mass of the perturbed scalar field, and $E_{\rm back}$ is the background scale and is manifestly smaller than the scale  $\Lambda_s$ at which perturbative unitarity breaks down. That scale is dominated by the cubic and quartic operators $\dot \varphi (\partial_i \varphi )^2$ and $(\p_i \varphi)^4$. For comparison $\mu_{500}$ represents the scale at which the operator $\varphi^5$ breaks tree-level unitarity. The example \eqref{eqn:PXsoln_direct} is specifically engineered so that the contribution of that operator and any other operator is sub-dominant.\vspace{-1cm}
 \label{fig:H}}
\end{figure}


For example, if we choose
\ba
&& \phi(t)=\phi_0 e^{\Lambda t} \, , \\
&&  h ( \Phi )= -  \frac{\Phi}{\phi_0} \frac{ 1 - \Phi^2/\phi_0^2 }{ 1 +\Phi^4/\phi_0^4 } \frac{ \Phi^{10}/\phi_0^{10} }{ 1 +\Phi^{20}/\phi_0^{20} }\,,
\label{eqn:h}
\ea
then the Hubble parameter goes through a bounce at $t=0$,
\ba
 H (t) = \frac{\Lambda^3}{2\mpl^2} \frac{\text{sinh} (\Lambda t) }{\text{cosh} (2 \Lambda t) \cosh(10 \Lambda t)} \, ,
\label{eqn:PXsoln_direct}
\ea
which is a smooth bounce that violates the NEC for a time $\Delta t \sim \Lambda^{-1}$ and produces a Hubble rate on the order of $H_m \sim \Lambda^{3} / \mpl^{2}$ on exiting the NEC violating region. This is shown in Figure \ref{fig:H}. To ensure that the perturbed field on that background is well-behaved, we can for instance choose,
\ba
q(\Phi)= \frac{\Lambda^2}{\mpl^2}  \(1+q_1 \frac{\phi_0^4/\Phi^4}{(1+\Phi/\phi_0 ) }  \)\,.
\ea
 As discussed in section~\ref{sec:highEnergyPX}, without a shift symmetry the model has to involve a hierarchy of scales and the ratio $\Lambda/\mpl$ is chosen to that effect (hence $H_m$, $q$, etc. are parametrically smaller than na\"{i}vely expected). This model would give the desired bouncing solution for $\phi (t), H(t)$, when  $\Lambda_c \to \infty$, however in that limit fluctuations on this background have negative sound speed (there is a gradient instability), and it badly violates unitarity. To remedy this, we switch on the high energy effects (\ie bring $\Lambda_c$ to a finite value) and include the irrelevant operator $(\Box \Phi)^2$ at that scale. As derived previously, this scale should be in the appropriate range to `save' unitarity, but without spoiling the low-energy effective theory.\\

For concreteness we plot the behaviour of the background and the strong coupling scale for the following specific choices of parameters,
\ba
\Lambda=10^{-3}\mpl\,,\quad \phi_0=0.5 \mpl\,,\quad q_1=6.4\quad {\rm and }\quad
\Lambda_c = 0.028 \mpl\,.
\ea
 For this choice of parameters,  we now have a finite positive sound speed and satisfy tree-level unitarity throughout the bounce. The strong coupling scale is indeed set by the cubic and quartic operator $\dot \varphi (\p_i\varphi)^2$ and $(\p_i\varphi)^4$, while the higher order effective interaction scales are all much larger than the strong coupling scale $\Lambda_s$. This example has indeed been specifically engineered so as to suppress the effect of any other operator.\\

\paragraph{Strong Coupling Scale:}
  An explicit calculation shows that within the NEC violating region, we have $A\sim 2$, $B\sim 2\times 10^{-3}$ and the strong coupling scale deduced from the $2-2$ scattering amplitude (including all the operators that would affect that amplitude) is $\Lambda_s\sim 10 \Lambda$. Then we can check that the mass of the fluctuations on that background is indeed small,
  \ba
  m^2_{\rm eff} \sim  \Lambda^2 \sim  10^{-2}\Lambda_s^2
  \ea
and the scale associated with the relevant operator $\varphi^3$ is sufficiently small, $\mu_{300}\sim  \Lambda$, as is the dimensionless coupling constant in front of the marginal operator $\varphi^4$, $\mu_{400}\sim \mathcal{O}(1)$. As for all the infinite number of irrelevant operators, their respective scales should be at least $\Lambda_s$. A direct calculation shows that within the region of interest (\ie within the NEC violating region), all the requirements derived from tree--level unitarity in (\ref{eq:req1}--\ref{eq:req4}) are indeed satisfied. For illustration purposes, we represent the scale $\mu_{500}$ we would have naively derived from where the quintic operator $\varphi^5$ breaks unitarity in Fig.~\ref{fig:H} and it indeed lies above $\Lambda_s$ throughout the bounce. The same remains true for all the other operators (other than $\dot \varphi (\p_i \varphi)^2$ and $(\p_i \varphi)^4$). \\

We can also directly see that the derivatives of the background remain small, obviously $\phi^{(n+1)}/\phi^{(n)} \sim \Lambda \ll \Lambda_s$ (the hierarchy involved in the example is not an important one, but it simply serves as an illustration of the principle). Moreover we can check that the variation of the mass and the coefficients of the kinetic matrix are small, $\dot B / B \sim 10^{-6}\mpl$, and $\dot A /A \sim 1.5 \times 10^{-3}\mpl$, which is actually what sets the scale of the background, $E_{\rm back}\sim 1.5\,  \Lambda$.  \\

We have therefore shown how, for couplings given by \eqref{eqn:PX_direct}, and solution \eqref{eqn:PXsoln_direct}, while if one took $\Lambda_c\to \infty$ the bounce would have a classical gradient instability and would badly violate perturbative unitarity, this can be remedied by including high energy effects that are present at  sufficiently high energy  (finite $\Lambda_c$) without severely affecting the predictions of the low energy EFT. For a bounce to occur without violating unitarity in a $P(\Phi,X)$ theory, the parameters have to be carefully tuned and a hierarchy of scale had to be introduced already in the $P(\Phi,X)$ model.

\section{Summary}
\label{sec:conc}

While solutions that violate the NEC condition are relatively easy to find classically, they may not be trusted if they are derived beyond the regime of validity of their effective field theory. In this work, we have derived the conditions set by tree-level unitarity on NEC violating effective field theories on a cosmological background. In $P(\Phi,X)$ theories minimally coupled to gravity, without including any higher energy effects, it is impossible to describe a NEC violation (much less a complete bounce): Any classical solution automatically severely violates unitarity and cannot be trusted. A natural resolution is to include the high energy effects, irrelevant operators suppressed by a higher energy scale that are naturally expected to be present. We have shown that these can regulate not only the classical instabilities that arise in these classical NEC violating solutions, but can also regulate scattering amplitudes, hence providing a much better handle on unitarity. We have derived the precise requirements set by tree-level unitarity in $P(\Phi,X)$ models with additional irrelevant operators, and shown that while the theory should be very carefully tuned, in principle there is a open window of possibility for a stable cosmological bounce that preserves unitarity. To further illustrate the constraints set by unitarity and level of tuning required, we have presented an explicit $P(\Phi,X)$ that generates a stable cosmological bounce and preserves unitarity within the region of the bounce, albeit at the price of introducing an unnaturally small parameter into, and hence finely tuning, the EFT. \\

The analysis and example provided here was not aimed at providing a full cosmological framework and we have only focused on the possibility of violating the NEC while avoiding any instability and remaining within the regime of validity of the effective field theory. We have not addressed the question of particle production throughout the bounce which is beyond the scope of this work, however given that there is no instability and the scales are under control, there is {\it a priori} no reason to expect a large particle production at this point.
Moreover, the high level of tuning required to obtain a stable and unitary bounce certainly raises the question of whether such a model would ever accommodate the precise cosmology we observe today and be able to reproduce the precise value of the spectral index (without an excess of non-Gaussianities, \cite{Gao:2014eaa}) and the tensor to scalar ratio to be embedded in a consistent and viable model for the cosmological history of our Universe.  \\

\noindent {\bf Acknowledgements:}
%
We wish to thank Anna Ijjas, Jean-Luc Lehners,  Paul Steinhardt and  Andrew Tolley  for very useful discussions and comments on the manuscript.
CdR is supported by a Royal Society Wolfson Merit Award. SM is supported by the von Clemm Fellowship and the Imperial College President's Scholarship.\vspace{1cm}

\appendix


\section{Bouncing with a Pure Ghost-Condensate}
\label{App:GC}

In this appendix we review the violation of the NEC in a pure ghost condensate model, \cite{ArkaniHamed:2003uy,Creminelli:2006xe,Buchbinder:2007ad,Koehn:2015vvy}.
Starting with the particular $P(X)$,
\ba
\label{eq:GhostCond}
P(X)=-p X + \frac{q}{\Lambda^4} X^2\,,
\ea
a violation of the NEC is possible if $p>0$. In that case the standard vacuum $\langle \phi \rangle = 0$ carries a ghost, but no ghost is present in  the `ghost-condensate phase' ($\langle \dot \phi \rangle \ne 0$) where the quadratic terms $X^2$ become relevant. Just like any other $P(\Phi, X)$ theory, this model is unstable and breaks unitarity even before entering the NEC violating region, unless higher energy effects are considered as discussed in section~\ref{sec:highEnergyPX} and so we have these high energy effects entering at $\Lambda_c$ in mind throughout this appendix. \\

The expression for the kinetic coefficients are
\ba
A = -p + \frac{6q X }{\Lambda^4} \quad{\rm and }\quad B = -p  + \frac{2q X }{\Lambda^4}\,,
\ea
so interestingly the variations of $A$ and $B$ are linked, $\dot A = 3\dot B$, and we have
\ba
X=\frac{\Lambda^4}{4q}(A-B)\,.
\ea
As a consequence, we therefore have
\ba
\label{eq:varX}
\frac{\dot X}{ X}=\frac{\dot A -\dot B }{A-B} = \frac{2 \dot B}{A-B}\quad\text{and similarly}\quad
\frac{\ddot X}{ X} = \frac{2 \ddot B}{A-B}\,.
\ea
Now from the Raychaudhuri equation, within the NEC violating region $\dot H>0$, we have
\ba
0< B < \mathcal{O}\(\frac{1}{\Lambda_c^2}\frac{\ddot X }{ X}\)\,,
\ea
where the exact expression on the right hand side depends on the very precise operators that enter at $\Lambda_c$ (and could also for instance involve terms of the form $H \dot X/ \Lambda_c^2 X $),  but this analysis is independent of the precise form (as it should). We merely use the fact that they involve higher derivatives as is required if those terms are to cure the instabilities associated with pure $P(X)$ bounce. Then from \eqref{eq:varX} within the NEC violating region we ought to have
\ba
 A-B < \mathcal{O}\(\frac{1}{\Lambda_c^2}\frac{\ddot B }{ B}\)\,.
\ea
Requiring that the background does not vary faster than $\Lambda_c'=\sqrt{A}\Lambda_c$ sets $\ddot B / B \ll A \Lambda_c^2$, which therefore implies that
\ba
 1-c^2_s \ll 1\,,
\ea
\ie the speed of sound should be very close to luminal. Now for a $P(X)$ model with close to luminal speed of sound the unitarity bound \eqref{eq:PXX} cannot be satisfied.

One may be worried that the bound \eqref{eq:PXX} is not technically valid if the speed of sound is not small since in going from \eqref{eq:mucFull} to \eqref{eq:mucSimp} we have assumed $A\gg B$. It is straightforward to rederive the bound \eqref{eq:PXX} when this assumption is relaxed and we then find that unitarity imposes
\ba
1\ll \Lambda_c^4 \bar P_{, XX} \ll \frac{(1-c_s^2)^4}{c_s}\,,
\ea
which is even more impossible to satisfy when $ 1-c^2_s \ll 1$. So we can conclude that a pure ghost-condensate model of the form \eqref{eq:GhostCond} can never give rise to a unitarity NEC violation (let alone a bounce) even when introducing higher derivative terms at a higher energy scale. The only way to avoid this argument in the ghost-condensate model is if the higher energy effects also involve operators that are not higher derivatives, \ie involve terms that break the shift symmetry at high energy. Instead of trying to maintain a low energy EFT that preserves the shift symmetry and only breaks that symmetry softly at high energy, in this manuscript we consider instead an explicit model that directly breaks the shift symmetry at low energy as is described in section~\ref{Explicitmodel}.


\newpage

\bibliographystyle{JHEP}
\bibliography{refs}

\providecommand{\href}[2]{#2}\begingroup\raggedright\begin{thebibliography}{10}

\bibitem{Barrow:1993hp}
J.~D. Barrow, {\it {Nonsingular scalar - tensor cosmologies}},  {\em Phys.
  Rev.} {\bf D48} (1993) 3592--3595.

\bibitem{Gasperini:1992em}
M.~Gasperini and G.~Veneziano, {\it {Pre - big bang in string cosmology}},
  {\em Astropart. Phys.} {\bf 1} (1993) 317--339,
  [\href{http://arxiv.org/abs/hep-th/9211021}{{\tt hep-th/9211021}}].

\bibitem{Khoury:2001wf}
J.~Khoury, B.~A. Ovrut, P.~J. Steinhardt, and N.~Turok, {\it {The Ekpyrotic
  universe: Colliding branes and the origin of the hot big bang}},  {\em Phys.
  Rev.} {\bf D64} (2001) 123522,
  [\href{http://arxiv.org/abs/hep-th/0103239}{{\tt hep-th/0103239}}].

\bibitem{Steinhardt:2001st}
P.~J. Steinhardt and N.~Turok, {\it {Cosmic evolution in a cyclic universe}},
  {\em Phys. Rev.} {\bf D65} (2002) 126003,
  [\href{http://arxiv.org/abs/hep-th/0111098}{{\tt hep-th/0111098}}].

\bibitem{Khoury:2001bz}
J.~Khoury, B.~A. Ovrut, N.~Seiberg, P.~J. Steinhardt, and N.~Turok, {\it {From
  big crunch to big bang}},  {\em Phys. Rev.} {\bf D65} (2002) 086007,
  [\href{http://arxiv.org/abs/hep-th/0108187}{{\tt hep-th/0108187}}].

\bibitem{ArkaniHamed:2003uy}
N.~Arkani-Hamed, H.-C. Cheng, M.~A. Luty, and S.~Mukohyama, {\it {Ghost
  condensation and a consistent infrared modification of gravity}},  {\em JHEP}
  {\bf 05} (2004) 074, [\href{http://arxiv.org/abs/hep-th/0312099}{{\tt
  hep-th/0312099}}].

\bibitem{Battefeld:2014uga}
D.~Battefeld and P.~Peter, {\it {A Critical Review of Classical Bouncing
  Cosmologies}},  {\em Phys. Rept.} {\bf 571} (2015) 1--66,
  [\href{http://arxiv.org/abs/1406.2790}{{\tt arXiv:1406.2790}}].

\bibitem{Peter:2006hx}
P.~Peter, E.~J.~C. Pinho, and N.~Pinto-Neto, {\it {A Non inflationary model
  with scale invariant cosmological perturbations}},  {\em Phys. Rev.} {\bf
  D75} (2007) 023516, [\href{http://arxiv.org/abs/hep-th/0610205}{{\tt
  hep-th/0610205}}].

\bibitem{Buchbinder:2007ad}
E.~I. Buchbinder, J.~Khoury, and B.~A. Ovrut, {\it {New Ekpyrotic cosmology}},
  {\em Phys. Rev.} {\bf D76} (2007) 123503,
  [\href{http://arxiv.org/abs/hep-th/0702154}{{\tt hep-th/0702154}}].

\bibitem{Peter:2008qz}
P.~Peter and N.~Pinto-Neto, {\it {Cosmology without inflation}},  {\em Phys.
  Rev.} {\bf D78} (2008) 063506, [\href{http://arxiv.org/abs/0809.2022}{{\tt
  arXiv:0809.2022}}].

\bibitem{Novello:2008ra}
M.~Novello and S.~E.~P. Bergliaffa, {\it {Bouncing Cosmologies}},  {\em Phys.
  Rept.} {\bf 463} (2008) 127--213, [\href{http://arxiv.org/abs/0802.1634}{{\tt
  arXiv:0802.1634}}].

\bibitem{lehners_ekpyrotic_2010}
J.-L. Lehners, {\it Ekpyrotic {Non}-{Gaussianity} -- {A} {Review}},  {\em
  Advances in Astronomy} {\bf 2010} (2010) 1--19. arXiv: 1001.3125.

\bibitem{Ijjas:2015zma}
A.~Ijjas and P.~J. Steinhardt, {\it {The anamorphic universe}},  {\em JCAP}
  {\bf 1510} (2015), no.~10 001, [\href{http://arxiv.org/abs/1507.03875}{{\tt
  arXiv:1507.03875}}].

\bibitem{brandenberger_bouncing_2016}
R.~Brandenberger and P.~Peter, {\it Bouncing {Cosmologies}: {Progress} and
  {Problems}},  {\em arXiv:1603.05834 [astro-ph, physics:gr-qc, physics:hep-ph,
  physics:hep-th]} (Mar., 2016). arXiv: 1603.05834.

\bibitem{Peter:2016kan}
P.~Peter and S.~D.~P. Vitenti, {\it {The simplest possible bouncing quantum
  cosmological model}},  {\em Mod. Phys. Lett.} {\bf A31} (2016), no.~21
  1640006, [\href{http://arxiv.org/abs/1603.02342}{{\tt arXiv:1603.02342}}].

\bibitem{Cai:2016thi}
Y.~Cai, Y.~Wan, H.-G. Li, T.~Qiu, and Y.-S. Piao, {\it {The Effective Field
  Theory of nonsingular cosmology}},  {\em JHEP} {\bf 01} (2017) 090,
  [\href{http://arxiv.org/abs/1610.03400}{{\tt arXiv:1610.03400}}].

\bibitem{Cai:2017tku}
Y.~Cai, H.-G. Li, T.~Qiu, and Y.-S. Piao, {\it {The Effective Field Theory of
  nonsingular cosmology: II}},  \href{http://arxiv.org/abs/1701.04330}{{\tt
  arXiv:1701.04330}}.

\bibitem{Visser:1999de}
M.~Visser and C.~Barcelo, {\it {Energy conditions and their cosmological
  implications}},  in {\em {Proceedings, 3rd International Conference on
  Particle Physics and the Early Universe (COSMO 1999): Trieste, Italy,
  September 27-October 3, 1999}}, pp.~98--112, 2000.
\newblock \href{http://arxiv.org/abs/gr-qc/0001099}{{\tt gr-qc/0001099}}.

\bibitem{Schulz:2001yx}
A.~E. Schulz and M.~J. White, {\it {The Tensor to scalar ratio of phantom dark
  energy models}},  {\em Phys. Rev.} {\bf D64} (2001) 043514,
  [\href{http://arxiv.org/abs/astro-ph/0104112}{{\tt astro-ph/0104112}}].

\bibitem{Abbott:1984qf}
L.~F. Abbott, {\it {A Mechanism for Reducing the Value of the Cosmological
  Constant}},  {\em Phys. Lett.} {\bf B150} (1985) 427--430.

\bibitem{Alberte:2016izw}
L.~Alberte, P.~Creminelli, A.~Khmelnitsky, D.~Pirtskhalava, and E.~Trincherini,
  {\it {Relaxing the Cosmological Constant: a Proof of Concept}},  {\em JHEP}
  {\bf 12} (2016) 022, [\href{http://arxiv.org/abs/1608.05715}{{\tt
  arXiv:1608.05715}}].

\bibitem{Hochberg:1997wp}
D.~Hochberg and M.~Visser, {\it {Geometric structure of the generic static
  traversable wormhole throat}},  {\em Phys. Rev.} {\bf D56} (1997) 4745--4755,
  [\href{http://arxiv.org/abs/gr-qc/9704082}{{\tt gr-qc/9704082}}].

\bibitem{Hochberg:1998ha}
D.~Hochberg and M.~Visser, {\it {Dynamic wormholes, anti-trapped surfaces, and
  energy conditions}},  {\em Phys. Rev.} {\bf D58} (1998) 044021,
  [\href{http://arxiv.org/abs/gr-qc/9802046}{{\tt gr-qc/9802046}}].

\bibitem{Hochberg:1998ii}
D.~Hochberg and M.~Visser, {\it {The Null energy condition in dynamic
  wormholes}},  {\em Phys. Rev. Lett.} {\bf 81} (1998) 746--749,
  [\href{http://arxiv.org/abs/gr-qc/9802048}{{\tt gr-qc/9802048}}].

\bibitem{Holdom:2004yx}
B.~Holdom, {\it {Accelerated expansion and the Goldstone ghost}},  {\em JHEP}
  {\bf 07} (2004) 063, [\href{http://arxiv.org/abs/hep-th/0404109}{{\tt
  hep-th/0404109}}].

\bibitem{Cline:2003gs}
J.~M. Cline, S.~Jeon, and G.~D. Moore, {\it {The Phantom menaced: Constraints
  on low-energy effective ghosts}},  {\em Phys. Rev.} {\bf D70} (2004) 043543,
  [\href{http://arxiv.org/abs/hep-ph/0311312}{{\tt hep-ph/0311312}}].

\bibitem{Dubovsky:2005xd}
S.~Dubovsky, T.~Gregoire, A.~Nicolis, and R.~Rattazzi, {\it {Null energy
  condition and superluminal propagation}},  {\em JHEP} {\bf 03} (2006) 025,
  [\href{http://arxiv.org/abs/hep-th/0512260}{{\tt hep-th/0512260}}].

\bibitem{Creminelli:2006xe}
P.~Creminelli, M.~A. Luty, A.~Nicolis, and L.~Senatore, {\it {Starting the
  Universe: Stable Violation of the Null Energy Condition and Non-standard
  Cosmologies}},  {\em JHEP} {\bf 12} (2006) 080,
  [\href{http://arxiv.org/abs/hep-th/0606090}{{\tt hep-th/0606090}}].

\bibitem{Deffayet:2010qz}
C.~Deffayet, O.~Pujolas, I.~Sawicki, and A.~Vikman, {\it {Imperfect Dark Energy
  from Kinetic Gravity Braiding}},  {\em JCAP} {\bf 1010} (2010) 026,
  [\href{http://arxiv.org/abs/1008.0048}{{\tt arXiv:1008.0048}}].

\bibitem{Pujolas:2011he}
O.~Pujolas, I.~Sawicki, and A.~Vikman, {\it {The Imperfect Fluid behind Kinetic
  Gravity Braiding}},  {\em JHEP} {\bf 11} (2011) 156,
  [\href{http://arxiv.org/abs/1103.5360}{{\tt arXiv:1103.5360}}].

\bibitem{Horndeski:1974wa}
G.~W. Horndeski, {\it {Second-order scalar-tensor field equations in a
  four-dimensional space}},  {\em Int. J. Theor. Phys.} {\bf 10} (1974)
  363--384.

\bibitem{Deffayet:2009mn}
C.~Deffayet, S.~Deser, and G.~Esposito-Farese, {\it {Generalized Galileons: All
  scalar models whose curved background extensions maintain second-order field
  equations and stress-tensors}},  {\em Phys. Rev.} {\bf D80} (2009) 064015,
  [\href{http://arxiv.org/abs/0906.1967}{{\tt arXiv:0906.1967}}].

\bibitem{Easson:2011zy}
D.~A. Easson, I.~Sawicki, and A.~Vikman, {\it {G-Bounce}},  {\em JCAP} {\bf
  1111} (2011) 021, [\href{http://arxiv.org/abs/1109.1047}{{\tt
  arXiv:1109.1047}}].

\bibitem{Rubakov:2013kaa}
V.~A. Rubakov, {\it {Consistent NEC-violation: towards creating a universe in
  the laboratory}},  {\em Phys. Rev.} {\bf D88} (2013) 044015,
  [\href{http://arxiv.org/abs/1305.2614}{{\tt arXiv:1305.2614}}].

\bibitem{Libanov:2016kfc}
M.~Libanov, S.~Mironov, and V.~Rubakov, {\it {Generalized Galileons:
  instabilities of bouncing and Genesis cosmologies and modified Genesis}},
  {\em JCAP} {\bf 1608} (2016), no.~08 037,
  [\href{http://arxiv.org/abs/1605.05992}{{\tt arXiv:1605.05992}}].

\bibitem{Kobayashi:2016xpl}
T.~Kobayashi, {\it {Generic instabilities of nonsingular cosmologies in
  Horndeski theory: A no-go theorem}},  {\em Phys. Rev.} {\bf D94} (2016),
  no.~4 043511, [\href{http://arxiv.org/abs/1606.05831}{{\tt
  arXiv:1606.05831}}].

\bibitem{Akama:2017jsa}
S.~Akama and T.~Kobayashi, {\it {Generalized multi-Galileons, covariantized new
  terms, and the no-go theorem for non-singular cosmologies}},
  \href{http://arxiv.org/abs/1701.02926}{{\tt arXiv:1701.02926}}.

\bibitem{Ijjas:2016tpn}
A.~Ijjas and P.~J. Steinhardt, {\it {Classically stable nonsingular
  cosmological bounces}},  {\em Phys. Rev. Lett.} {\bf 117} (2016), no.~12
  121304, [\href{http://arxiv.org/abs/1606.08880}{{\tt arXiv:1606.08880}}].

\bibitem{Ijjas:2016vtq}
A.~Ijjas and P.~J. Steinhardt, {\it {Fully stable cosmological solutions with a
  non-singular classical bounce}},  {\em Phys. Lett.} {\bf B764} (2017)
  289--294, [\href{http://arxiv.org/abs/1609.01253}{{\tt arXiv:1609.01253}}].

\bibitem{Koehn:2015vvy}
M.~Koehn, J.-L. Lehners, and B.~Ovrut, {\it {Nonsingular bouncing cosmology:
  Consistency of the effective description}},  {\em Phys. Rev.} {\bf D93}
  (2016), no.~10 103501, [\href{http://arxiv.org/abs/1512.03807}{{\tt
  arXiv:1512.03807}}].

\bibitem{Vikman:2004dc}
A.~Vikman, {\it {Can dark energy evolve to the phantom?}},  {\em Phys. Rev.}
  {\bf D71} (2005) 023515, [\href{http://arxiv.org/abs/astro-ph/0407107}{{\tt
  astro-ph/0407107}}].

\bibitem{Easson:2016klq}
D.~A. Easson and A.~Vikman, {\it {The Phantom of the New Oscillatory
  Cosmological Phase}},  \href{http://arxiv.org/abs/1607.00996}{{\tt
  arXiv:1607.00996}}.

\bibitem{Battarra:2014tga}
L.~Battarra, M.~Koehn, J.-L. Lehners, and B.~A. Ovrut, {\it {Cosmological
  Perturbations Through a Non-Singular Ghost-Condensate/Galileon Bounce}},
  {\em JCAP} {\bf 1407} (2014) 007, [\href{http://arxiv.org/abs/1404.5067}{{\tt
  arXiv:1404.5067}}].

\bibitem{Burgess:2007pt}
C.~P. Burgess, {\it {Introduction to Effective Field Theory}},  {\em Ann. Rev.
  Nucl. Part. Sci.} {\bf 57} (2007) 329--362,
  [\href{http://arxiv.org/abs/hep-th/0701053}{{\tt hep-th/0701053}}].

\bibitem{Creminelli:2010qf}
P.~Creminelli, G.~D'Amico, M.~Musso, J.~Norena, and E.~Trincherini, {\it
  {Galilean symmetry in the effective theory of inflation: new shapes of
  non-Gaussianity}},  {\em JCAP} {\bf 1102} (2011) 006,
  [\href{http://arxiv.org/abs/1011.3004}{{\tt arXiv:1011.3004}}].

\bibitem{deRham:2014fha}
C.~de~Rham, L.~Heisenberg, and R.~H. Ribeiro, {\it {Ghosts and matter couplings
  in massive gravity, bigravity and multigravity}},  {\em Phys. Rev.} {\bf D90}
  (2014) 124042, [\href{http://arxiv.org/abs/1409.3834}{{\tt
  arXiv:1409.3834}}].

\bibitem{Aydemir:2012nz}
U.~Aydemir, M.~M. Anber, and J.~F. Donoghue, {\it {Self-healing of unitarity in
  effective field theories and the onset of new physics}},  {\em Phys. Rev.}
  {\bf D86} (2012) 014025, [\href{http://arxiv.org/abs/1203.5153}{{\tt
  arXiv:1203.5153}}].

\bibitem{deRham:2014wfa}
C.~de~Rham and R.~H. Ribeiro, {\it {Riding on irrelevant operators}},  {\em
  JCAP} {\bf 1411} (2014), no.~11 016,
  [\href{http://arxiv.org/abs/1405.5213}{{\tt arXiv:1405.5213}}].

\bibitem{Gao:2014eaa}
X.~Gao, M.~Lilley, and P.~Peter, {\it {Non-Gaussianity excess problem in
  classical bouncing cosmologies}},  {\em Phys. Rev.} {\bf D91} (2015), no.~2
  023516, [\href{http://arxiv.org/abs/1406.4119}{{\tt arXiv:1406.4119}}].

\end{thebibliography}\endgroup

\end{document}